\documentclass[manuscript=article]{achemso}
\usepackage[version=3]{mhchem} 
\usepackage{braket}
\usepackage{bbold}
\usepackage{subcaption}
\usepackage{tikz}
\usepackage{amsmath}

\SectionNumbersOn

\author{Brieuc Le D\'e}
\author{Simon Huppert}
\affiliation[INSP]
{Sorbonne Universit\'e, CNRS, Institut des NanoSciences de Paris, 4 place Jussieu, 75005 Paris, France}%
\author{Riccardo Spezia}
\affiliation[LCT]
{Sorbonne Universit\'e, CNRS, Laboratoire de Chimie Th\'eorique, 4 place Jussieu, 75005 Paris, France}%
\author{Alex W. Chin}
\email{alex.chin@insp.upmc.fr}
\affiliation[INSP]
{Sorbonne Universit\'e, CNRS, Institut des NanoSciences de Paris, 4 place Jussieu, 75005 Paris, France}%

\title[Extending Non-Perturbative Simulation Techniques for Open-Quantum Systems to Excited-State Proton Transfer and Ultrafast Non-Adiabatic Dynamics]
{Extending Non-Perturbative Simulation Techniques for Open-Quantum Systems to Excited-State Proton Transfer and Ultrafast Non-Adiabatic Dynamics}

\abbreviations{MPSDynamics}

\begin{document}


\begin{tocentry}
\includegraphics[scale=.16]{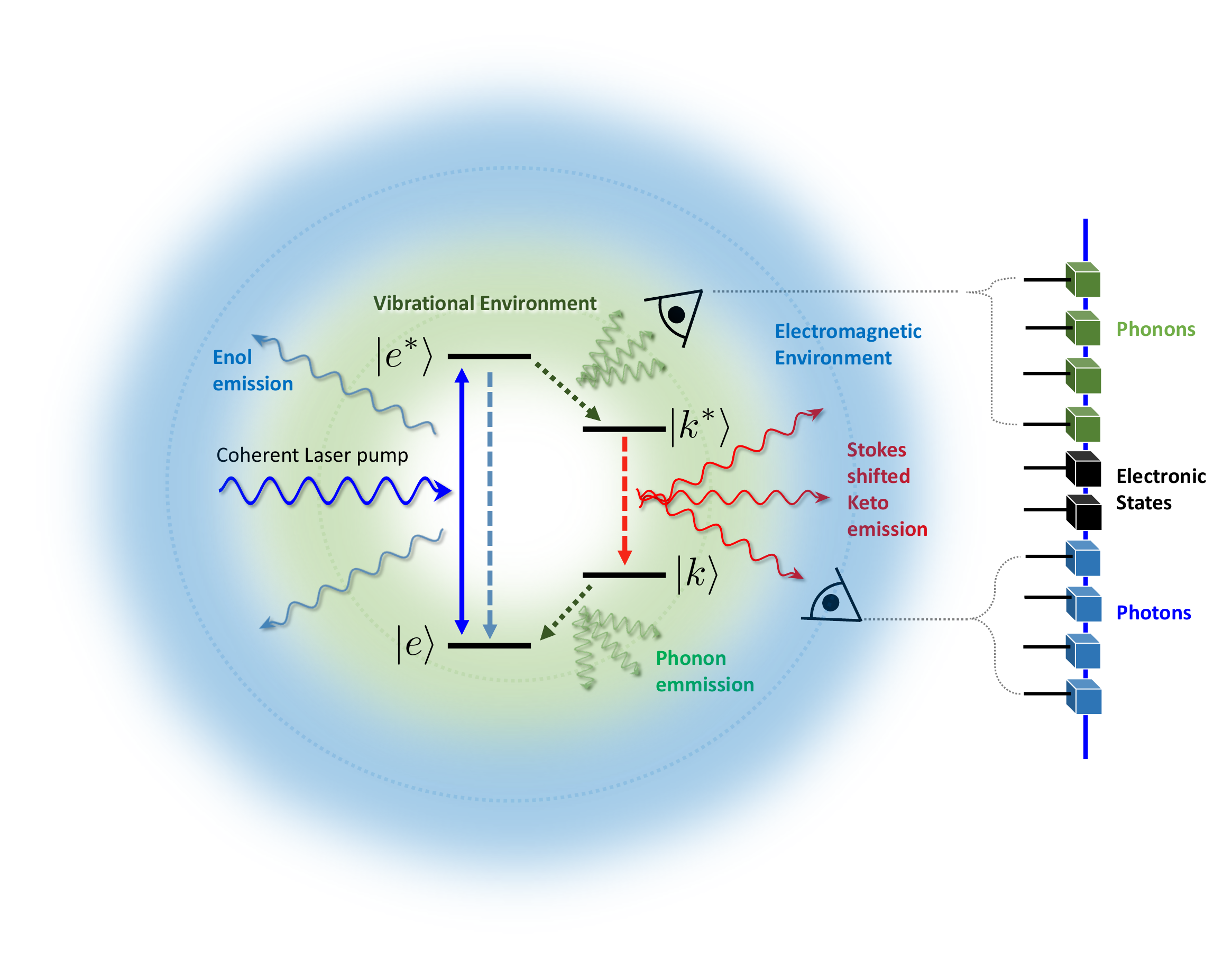}
\end{tocentry}

\begin{abstract}
Excited state proton transfer is an ubiquitous phenomenon in biology and chemistry, spanning from the ultrafast reactions of photo-bases and acids to light-driven, enzymatic catalysis and photosynthesis. However, the simulation of such dynamics involves multiple challenges, since high-dimensional, out-of-equilibrium vibronic states play a crucial role, while a fully quantum description of the proton's dissipative, real-space dynamics is also required. In this work, we extend the powerful Matrix Product State approach to open quantum systems (TEDOPA) to study these demanding dynamics, and also more general non-adiabatic processes that can appear in complex photochemistry subject to strong laser driving. As an illustration, we initially consider an open model of a four-level electronic system interacting with hundreds of intramolecular vibrations that drive ultrafast excited state proton transfer, as well as an explicit photonic environment that allows us to directly monitor the resulting dual fluorescence in this system.  We then demonstrate how to include a continuous 'reaction coordinate' of the proton transfer that allows numerically exact simulations that can be understood, visualized and interpreted in the familiar language of diabatic and adiabatic dynamics on potential surfaces, while also retaining an exact quantum treatment of dissipation and driving effects that could be used to study diverse problems in ultrafast photochemistry.
\end{abstract}


\section{Introduction}
Proton transfer is an elementary physical process that underlies a vast number of chemical transformations which play critical roles in biology \cite{reece_proton-coupled_2009,lama_unraveling_2022}, catalysis, and environmental energy harvesting \cite{ren2018molecular,lennox2017excited}. Excited-State Intramolecular Proton Transfer (ESIPT) is a more specific transfer that is triggered by optical and/or electronic excitation \cite{joshi_excited-state_2021}. This type of proton transfer, which can occur on timescales spanning $10^1-10^3$ fs \cite{joshi_excited-state_2021}, has garnered attention in the last 20 years since it holds several interesting characteristics. For example, the extremely large Stokes-shift in optical emission following ESIPT in some molecules ($\ge 8000~ \text{cm}^{-1}$) could make it an efficient fluorescent sensor for optical sensing \cite{chen_excitedstate_2021,sedgwick_excited-state_2018, Li2021} and biological imaging \cite{Reece2009,santos2021design}; and the related dual-colour emission from some ESIPT systems have been proposed for white-light generation in solid-state (organic) LEDS \cite{joshi_excited-state_2021}. ESIPT is also of interest, as it allows transport and storage of excitations in energy conversion processes \cite{goyal_tuning_2017, joshi_excited-state_2021}, light-driven enzymatic and artificial enzyme biosynthesis \cite{Liang2019}, and may provide a way to control ultrafast energy transfer in the nascent field of polaritonic chemistry \cite{Sokolovskii2024}. A comprehensive list of other possible applications can be found in a recent review by Joshi \textit{et al}. \cite{joshi_excited-state_2021}.     

Although fundamental, ESIPT is a challenging mechanism to study, both experimentally and theoretically \cite{jankowska_modern_2021,hammes-schiffer_proton-coupled_2012}. The hydrogen being a light nucleus, proton transfer can occur on an ultrafast timescale that is at the limit of many experimental probes, and the identification of the key 'reaction coordinates' for ESIPT rapidly becomes problematic for increasing molecule sizes, with some systems suggested to undergo ESIPT under the control of the molecular 'skeletal' vibrations\cite{Pijeau2017, kim_non-bornoppenheimer_2020, picconi_nonadiabatic_2021}, or directly along through-space, hydrogen-bond coordinates that could themselves be multidimensional, in the presence of protic solvent or protein environments \cite{xu_first-principles_2023,schnorr_direct_2023}. ESIPT, and more generally proton transfer, can also be highly correlated with electron transfer, leading to proton-coupled electron transfer (PCET) \cite{weinberg2012proton,lennox2017excited}. This is a rich and important field in its own right, but lies beyond the scope of the present article.


Given the wide dynamical range of typical molecular, solvent and host-material vibrations, ultrafast ESIPT will manifestly involve a number of out-of-equilibrium vibronic states whose dynamics and fluctuations will be temporally entangled with the dissipative (irreversible) dynamics of ESIPT. These lead to strong 'memory'-like effects that can cause the fate of the process to depend on the way the system was initially \emph{prepared}. Indeed, while describing strongly coupled vibrational 'environments' always entails the consideration of a large number of degrees of freedom, here, the situation is made more challenging by the fact that no single dynamical approximation, such as the Markov approximation, can be invoked for the entire bath: the proton-environment behaves as a quantum many body system whose 'reaction' coordinates may be time-dependent and/or collective/hybrid/emergent modes of the 'environmental' degrees of freedom. However, a number of time-resolved  optical spectroscopy techniques can now potentially probe such details of ESIPT dynamics \cite{lee_active_2013,schriever_interplay_2008,kim_coherent_2009,lochbrunner_microscopic_2003, takeuchi_coherent_2005, Wang2021}, including X-ray methods that also report on real-time changes of molecular structure during ESIPT \cite{Yin2023}. Predicting and understanding the underlying photodynamics of ESIPT at a numerically exact level of theory -- limited only by the quality of model parameterization -- is thus a timely problem.


Previous studies of ESIPT have used a variety of theoretical methods to treat particular aspects of this problem at different levels of approximation. Approaches starting from first principles, and which highlight the role of the molecular and heterogeneous environment on ESIPT, range from QM/MM free energy simulations invoking semi-classical (adiabatic) approximations, to simulations that predict non-adiabatic reaction rates \cite{zhao_excited_2021,zhong_probing_2024}. Recently, a promising \textit{ab initio} approach that blends real-time propagation of TD-DFT with quantum nuclear-electronic orbital dynamics has been presented and applied to ESIPT in condensed-phase environments \cite{yu_nonadiabatic_2022,xu_first-principles_2023,Chow2023}. When potential energy surfaces are available, other methods of treating the quantum nature of the proton dynamics and non-adiabaticity include semi-classical trajectory methods such as Surface Hopping \cite{Toldo2023}, or Exact Factorisation \cite{Agostini2019}; and for model vibronic systems, exact quantum dynamics have also been obtained by Multi-configurational Time-Dependent Hartree techniques \cite{Perveaux2017,Anand2020}. An excellent review of these  -- and other -- theoretical chemistry approaches to ESIPT has been given by Jankowska \textit{et al}. \cite{jankowska_modern_2021}.



In this article we attempt another fully quantum mechanical approach, based on the idea that ESIPT can be considered as a complex type of open quantum system  \cite{breuer2002theory,zhang2022modeling}. Indeed, a recent study of ESIPT in full-molecule-names of HBQ, HBT by Zhang \textit{et al.} examined the non-adiabiatic ESIPT dynamics in a diabatic picture of electronic-proton states coupled via a $1D$ reaction coordinate in the presence of a bath representing the rest of the molecule vibrations \cite{zhang2022modeling}. By treating the environment as a simple Markovian bath, they derived a Markovian Lindblad master equation for ESIPT, including the explicit role of the laser excitation, which is problematic for many of the methods mentioned above, such as the commonly used surface hopping methods. Inspired by this viewpoint, here we extend the numerically exact Time-Evolving Density with Orthogonal Polynomial (TEDOPA) method for Non-Markovian open systems to the problem of ESIPT.              

TEDOPA has been established over the last decade as a state-of-the-art method for treating the coupling of a quantum system to its environmental excitations in a range of condensed matter, quantum optical and molecular problems, and is particularly efficient in the case of linear-vibronic models \cite{Prior2010,prior2013quantum,chin2013role,del2018tensor,schroder2019tensor,dunnett2021influence,hunter2024environmentally}. At the heart of the method is the efficient representation of the evolving system-environment wave function in terms of either Matrix Product States (MPS) or Tensor Network States (TNS) \cite{banuls_tensor_2023}, with rather interesting connections to the tree-structures found in ML-MCTDH \cite{larsson2024tensor}. The essential properties of these states and why these are particularly powerful for open system problems are presented in Section \ref{sec:MPS}, but a key feature is the access to the complete quantum state of the system \emph{and} its environments. Unlike master equations in which the environment has been removed from the problem (traced out), TEDOPA allows one to track and visualize the environmental dynamics leading to dissipative phenomena in the system \cite{del2018tensor,schroder2019tensor,lacroix2024non}.

This capability could be particularly useful for photo-chemistry problems, as all modes of the environment are treated on an equal (exact) footing, and interrogation of the environment state can provide ways to identify relevant reaction coordinates on the fly (important in multi-stage dynamics), and/or visualize high dimensional dynamics in a purely diabatic basis \cite{schroder2019tensor,hunter2024environmentally}. 
Indeed, the use of TNS and MPS wave functions, also known as tensor-trains in the applied mathematics literature \cite{oseledets2011tensor,lubich2015time}, for simulating high-dimensional excited-state dynamics has seen tremendous growth over the last few years \cite{borrelli2021finite,paeckel_time-evolution_2019,lyu2022tensor,peng2023studies,greene2017tensor,mangaud2023survey}.

However, previous work employing TEDOPA has typically explored how environments drive transitions between discrete states of point-like systems, although it is capable of treating spatial-temporal correlations between point-like systems embedded in different regions of a common bath \cite{lacroix2024non}. For ESIPT systems possessing well-defined reaction coordinates, describing dissipative dynamics requires an extension of the current algorithms to treat a system evolving on one or more \emph{continuous} spatial degrees of freedom. This article presents our first efforts in this direction, where we use toy models similar to those of Zhang \textit{et al}. to demonstrate that TEDOPA can be used to visualize dissipative dynamics of ESIPT in real space with an exact treatment of the vibrational environment, the coupling to the electromagnetic environment that leads to dual fluorescence, as well as driving by laser pulses of arbitrary strength.                

In this work, we progressively present these capabilities. After introducing the MPS framework in Section~\ref{sec:MPS}, we start from a purely open system model without reaction coordinates  (Section~\ref{sec:4lvl}) to show how access to environmental information allows us to correlate, for example, the real-time emergence of dual fluorescence with the energy dissipated into the vibrational environment. This model also allows us to explore the role of strong and weak driving on the emission properties and dynamics of ESIPT, anticipating future work on how polaritonic effects could also alter ESIPT phenomena. We then introduce an explicit reaction coordinate in Section~\ref{sec:RC} and present examples where we are able to give a complete spatio-temporal description of ESIPT, including dissipation and an exact treatment of decoherence in the non-adiabatic dynamics, which often has to be treated in an \textit{ad hoc} fashion in surface hopping methods \cite{shu2023decoherence}. Importantly, we also introduce in Section~\ref{subsec:adiabproj}  a matrix product operator projector that allows to re-express the dynamics in the chemically familiar language of adiabatic potential surfaces, allowing for future comparisons with other more established formalisms, and which also provides a quantitative measure of the 'non-adiabaticity' of the dynamics. We then summarise our results in Section~\ref{sec:conclusion}. Overall, this yields a versatile MPS/TNS framework designed for the study of excited-state chemical problems, as well as the study of non-adiabatic transitions, which we hope will be useful for researchers working in these areas.

\section{A Brief Introduction to Matrix Product States}
\label{sec:MPS}

The calculations in this work make use of the Matrix Product State (MPS) Ansatz for many-body quantum states \cite{banuls_tensor_2023,cirac_matrix_2021,orus_practical_2014}. A general N-body wave function $\ket{\psi}$ can be written on the basis of, say, $d$ one-body substates $\{\ket{\phi_{i}} \}$ as $\ket{\psi} = \sum_{\{i\}} C_{i_1\ i_2 \ldots i_N}\ket{\phi_{i_1}\ldots\phi_{i_N}}$. The coefficients $C_{i_1\ i_2 \ldots i_N}$ are the amplitudes of each N-body configuration to the full system state. Clearly, the total number of amplitudes scales as $d^N$, leading to the ‘curse of dimensionality' that rapidly makes any direct numerical treatment extremely expensive. 

However, starting from a well-defined initial configuration, most physical Hamiltonians will only allow a much smaller sub-space of states to be explored in a finite time. Additionally, if the initial state is a product state, then it seems reasonable that the size of this subspace can be \emph{classified} by the amount of the quantum correlation (entanglement) that is generated during the dynamics. Indeed, in the absence of any entanglement, the problem would reduce to a type of mean-field description that would have linear scaling in $N$.  

\begin{figure}
    \centering
        \includegraphics[scale=1.0]{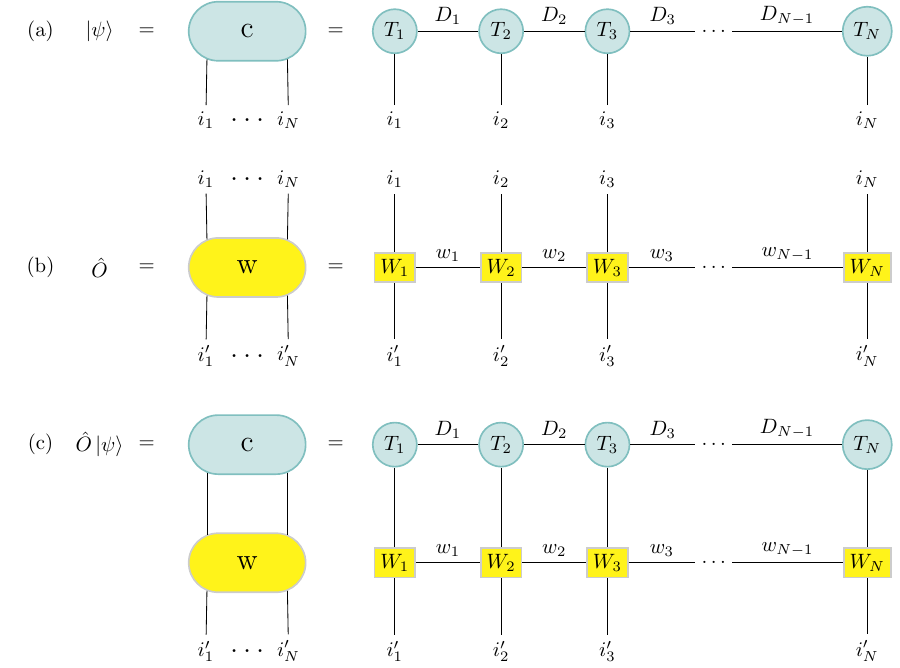}
    \caption{Diagram representing the Matrix Product State Ansatz. (a) The full wave-function represented by the multidimensional coefficient $C_{i_1\ i_2 \ldots i_N}$ can be approximated by dividing it into smaller tensors. The $i_1 \ldots i_N$ coefficients are physical dimensions while $D_1 \ldots D_N$ are bond dimensions, i.e. entanglement between tensors. (b) The MPO $\hat{O}$ is composed of local tensors $W_j$ connected by bond dimensions $w_j$. It has twice the number of physical dimension to allow $\hat{O} \ket{\psi}$ and $ \bra{\psi} \hat{O}$ operations.(c) $W_j$ MPO tensors are acting on MPS tensors $T_j$. Common physical dimensions between the MPS and the MPO are merged to lead to a new wave-function. Legs of dimension one at the extremities of the chains are not represented in the diagrams.} \label{fig:MPS_tensor_intro}
\end{figure}

Matrix Products States are essentially low-rank decompositions of the full wave function that are formed by breaking up the multi-linear array (a tensor) $C_{i_1\ i_2 \ldots i_N}$  into a product of smaller data arrays associated with each particle \cite{schollwock2011density,banuls_tensor_2023}. The MPS decomposition can be represented with the help of diagrams as in Figure~\ref{fig:MPS_tensor_intro}. In this formalism, the tensor $C_{i_1\ i_2 \ldots i_N}$ is written as several smaller tensors $T$ (Figure~\ref{fig:MPS_tensor_intro} (a)). Each tensor $T$ has three legs (array dimensions), with legs pointing towards the bottom of the diagram ($i_j$, $j \in [1,N]$) running over physical dimension of the one-particle states on that site (for example the Fock dimension or the number of electronic states). The legs linking tensors represent the entanglement between tensors $T_j$ and $T_{j+1}$, and contain $D_j$ parameters, known as the local ‘bond dimension' that allow the MPS to describe non-classical correlated states ($D=1$ would be a product state).   

Tensors sharing a common leg implies contraction (Einstein summation, or equivalently a trace) over the shared dimensions of their arrays. Thus for a particular physical state configuration, i.e. if the elements of each downward pointing leg are specified, it can be seen that the scalar coefficient $C_{i_1\ i_2 \ldots i_N}$ is obtained from contractions of all the tensors in the state, which in $1D$ corresponds to matrix multiplication. The key aspect of the MPS Ansatz is that the number of parameters required to describe the state is now $N\times d\times D^2$, which has linear scaling with $N$, provided that $D$ is not itself a function of $N$. Importantly, rigorous results concerning  the so-called 'area laws' show that this is indeed the case for gapped $1D$ Hamitonians with short-range interactions \cite{schollwock2011density}, so that the $\{D_j\}$ can be used as convergence parameters for numerical simulations of such models \cite{banuls_tensor_2023, schollwock2011density}. Crucially for the dynamics of open quantum systems, it is possible to prove that many common system-environment Hamiltonians are formally equivalent to $1D$ models with  nearest-neighbour couplings, and that these models can be easily constructed via a unitary transformation of the environmental degrees of freedom \cite{Tamascelli2020,Prior2010, chin_exact_2010}. This is the underlying reason for why MPS are particularly useful for system-environment problems, and an example of the TEDOPA workflow, i.e. model$\rightarrow$chain mapping$\rightarrow$MPS simulation is given in Section \ref{sec:4lvl}.      
For the results presented in this paper, we typically find converged results for chains with $D\approx 10$, $d= 10$ and number of modes going from $N= 150$ to $N= 540$. Only for the less demanding simulation, it results to a total of $N \times d \times D^2 = 15 \times 10^4$ parameters in comparison with the full-rank wave function that would have $d^N = 10^{150}$ elements, making the numerical cost reduction clear. For more details about parameters, see Tables (S1-S7) in the Supporting Information.

Within the same formalism, operators can also be expressed as Matrix Product Operators (MPOs) by writing them as a chain of local tensors $\hat{O} = \sum_{\{i\}, \{i^\prime\}} W_{i^\prime_1\ i^\prime_2 \ldots i^\prime_N}^{i_1\ i_2 \ldots i_N}\ket{\phi_{i_1}\ldots\phi_{i_N}} \bra{\phi_{i^\prime_1}\ldots\phi_{i^\prime_N}}$ (Figure~\ref{fig:MPS_tensor_intro} (b)) \cite{strathearn_efficient_2018}. Accordingly, an MPO can act on an MPS to lead to a new wave-function (Figure~\ref{fig:MPS_tensor_intro} (c)). Effectively, the MPO form allows us to measure observables of the system -- \emph{or} the environment -- as the action of a local operator acting on the states (tensor) of a given site, which can be combined with the gauge freedom of the MPS representation to create a highly efficient procedure for evaluating multi-partite observables \cite{orus_practical_2014,schollwock2011density}. 

Importantly, this also applies to the Hamiltonian operator in MPO form, allowing for efficient time-evolution of the state by sweeping through a sequence of updates that act on no more than one or two site tensors at a time \cite{schollwock2011density,haegeman_unifying_2016,banuls_tensor_2023}. For (bosonic) open system problems, a particularly efficient and versatile evolution method has been shown to be the one-site \textit{Time Dependent Variational Principal} (1TDVP) and its recent extensions to allow for dynamic updates to the size of the bond dimensions \cite{paeckel_time-evolution_2019,haegeman_unifying_2016,dunnett2021efficient,garcia-ripoll_time_2006,yang2020time,rams2020breaking}. Further details about TDVP and its algorithmic implementation can be found in Ref \cite{dunnett2021efficient}. 
In this article, all of the TEDOPA calculations are performed using our source-source Julia package MPSDynamics.jl \cite{MPSDynamics}. This package, along with extensive documentation, tutorial examples and all of the models that we have been studying for research problems in physics and chemistry can be found at  \url{https://github.com/shareloqs/MPSDynamics}. We now return to the particular application for this article: ESIPT. 

\section{A Four-level Open-System Description of ESIPT }
\label{sec:4lvl}

Excited-state intramolecular proton transfer (ESIPT) is often described as a four-state cycle, representing the enol electronic ground state $\ket{e}$, the enol electronic excited state $\ket{e^*}$ and their keto counterparts ($\ket{k}$, $\ket{k^*}$), with the ESIPT process corresponding to the $\ket{e^*} \rightarrow \ket{k^*}$ transfer. In order to study an excited electronic transfer as well as emission from these states, the full four electronic-state cycle is simulated. Interactions with environments can be included along with the main system description. The coupling between a system and a discrete set of modes can be efficiently implemented within the MPS approach, without any loss of information. These modes can describe different environments such as molecular vibrations and photons. In fact, photons are treated in a similar manner as vibrations: both are represented by a set of discrete harmonic modes described by the associated (bosonic) creation and annihilation operators. Such environments are characterised by their spectral density, specifying the interaction between the system and each mode of the environment. The quantum dynamics of the open system wavefunction (including the environments) allows to capture the effect of the motion of the molecule on the electronic population dynamics as well as photo-emission and absorption processes.


\begin{figure}
    \centering
    \includegraphics[scale=1.0]{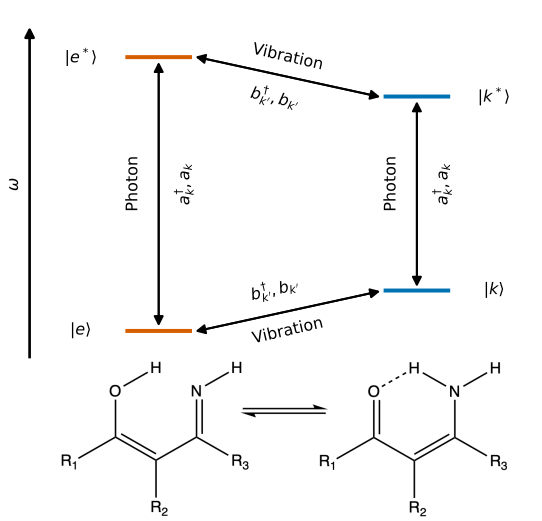}
    \caption{Four-state cycle that illustrates ESIPT. Interaction with environments allows transitions between states. Transitions from excited state to ground state correspond to photon coupling ($a_k^\dagger , a_k$ in eq~\ref{eq:4lvl}-\ref{eq:4lvl_end}) whereas transitions between excited states or ground states are due to vibration coupling ($b_{k^\prime}^\dagger , b_{k^\prime}$ in eq~\ref{eq:4lvl}-\ref{eq:4lvl_end}). The transfer $\ket{e^*} \rightarrow \ket{k^*}$ represents the excited-state proton transfer. A general ESIPT compound is shown on the bottom with $R_1$, $R_2$ and $R_3$ as general moities.}
    \label{fig:4lvl_scheme}
\end{figure}

To illustrate the effect of these couplings, the prototype system of Figure~\ref{fig:4lvl_scheme} is studied. In this model, there is no direct coupling between electronic states. Instead, environments mediate transitions between states: intramolecular vibrations allow to switch from one electronic structure to another and photons allow vertical transitions.

\subsection{Hamiltonian of Electrons Interacting with Photons and Vibrations}

To represent a four-state cycle interacting with environments (Figure~\ref{fig:4lvl_scheme}), a Hamiltonian can be written to account for these couplings. This Hamiltonian is composed of a system part $H_\text{S}$, a time-dependent term modelling the effect of an external driving field $H_\text{drive}(t)$, a bath component $H_\text{B}$ and an interaction term $H_\text{SB}$. The full Hamiltonian reads (here and in the following, we use arbitrary units in which $\hbar = 1$): 
\begin{subequations}
\renewcommand{\theequation}{\theparentequation.\arabic{equation}}
\begin{align} \label{eq:4lvl}
&H\left( t \right) = H_\text{S}  + H_\text{drive}\left( t \right) + H_\text{B} + H_\text{SB} \\ 
    &H_\text{S} = \omega_{e} \ket{e}\bra{e} + \omega_{e^*} \ket{e^*}\bra{e^*} + \omega_{k} \ket{k}\bra{k} + \omega_{k^*} \ket{k^*}\bra{k^*} \\
    &H_\text{drive}\left( t \right) = \left( \ket{e^*}\bra{e} + \ket{k^*}\bra{k}  + \text{h.c.} \right) \epsilon (t) \sin\left( \omega_\text{drive} t \right) \label{eq:4lvl_hst} \\
    &H_\text{B} = \overbrace{\sum_k \omega_k^\text{phot} \left(a_k^{\dagger}a_k\right)}^{\text{Photons}} + \overbrace{\sum_{k^\prime} \omega_{k^\prime}^\text{vib} \left(b_{k^\prime}^{\dagger}b_{k^\prime}\right)}^{\text{Vibrations}}  \\
    &H_\text{SB} =  \left(\ket{e^*}\bra{e} + \ket{k^*}\bra{k} + \text{h.c} \right) \sum_k g_k \left(a_k^{\dagger} + a_k \right)  + \left(\ket{e^*}\bra{k^*} + \text{h.c} \right) \sum_{k^\prime} y_{k^\prime} \left( b_{k^\prime}^{\dagger} +  b_{k^\prime} \right) \label{eq:4lvl_end}
\end{align}
\end{subequations}

\noindent with $a_k^{\left(\dagger\right)}$ the annihilation (creation) operator for the $k^\text{th}$ mode in the photon bath of frequency $\omega_k$, and $b_{k^\prime}^{\left(\dagger\right)}$ the annihilation (creation) operator for the $k^{^\prime \text{th}}$ mode in the vibration bath of frequency $\omega_{k^\prime}$. 
$\omega_e$, $\omega_{e^*}$,$\omega_k$ and $\omega_{k^*}$ are respectively the energy of the enol ground state, enol excited state, keto ground state and keto excited state. We also define $\Delta \omega_e = \omega_{e^*} - \omega_e$ and $\Delta \omega_k = \omega_{k^*} - \omega_k$. The coefficients $g_k$ and $y_{k^\prime}$ are determined for the different $k$ and $k^\prime$ indices defining the weights of the spectral density characterising the coupling between environments and the system for every mode. The spectral densities read $J_\text{phot}\left(\omega\right)=\sum_k g_k^2 \delta\left(\omega_k - \omega \right)$ and $J_\text{vib}\left(\omega\right)=\sum_k y_k^2 \delta\left(\omega_k - \omega \right)$. The time-dependent driving field is characterised by its frequency $\omega_\text{drive}$ and its (possibly time-dependent) amplitude $\epsilon (t) $.\\

The vibration bath couples the states $\ket{e^*}$ and $\ket{k^*}$ as well as $\ket{e}$ and $\ket{k}$ whereas photons couple the excited states with their respective ground states. The transfer between $\ket{e^*}$ and $ \ket{k^*}$ is the excited state proton transfer and the transfer between $\ket{k}$ and $ \ket{e}$ corresponds to ground state proton transfer. In this particular example, the ground state proton transfer is ignored.

\begin{figure}
    \centering
        \includegraphics[scale=1.0]{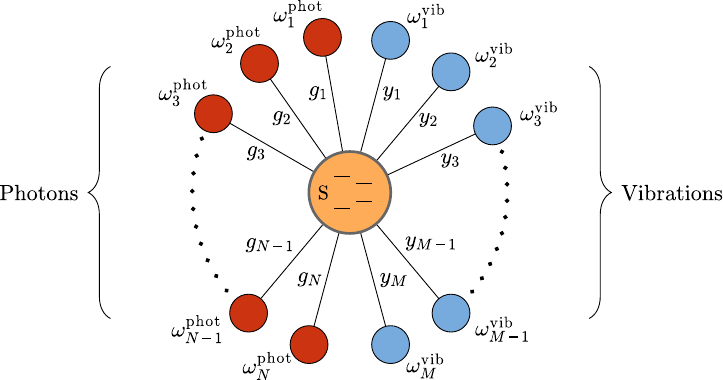}
    \caption{Star conformation. The electronic system (orange) is interacting with two different environments : $N$ photon modes (red) characterised by the spectral density $J_\text{phot}\left(\omega\right)=\sum_{k=1}^N g_k^2 \delta\left(\omega_k - \omega \right)$ on one side and with $M$ vibration modes (blue) characterised by the spectral density $J_\text{vib}\left(\omega\right)=\sum_{k=1}^M y_k^2 \delta\left(\omega_k - \omega \right)$ on the other side.}
    \label{fig:4lvl_star}
\end{figure}

This Hamiltonian corresponds to an electronic system interacting with several vibration modes of different frequencies. It is represented in Figure~\ref{fig:4lvl_star} in the form known as the ‘star-like' conformation, where each vibration mode is coupled schematically to the system situated at the centre. 
The star conformation is not directly adapted to the MPS formalism that naturally deals with chain-like systems in which each environment mode interacts with only its nearest neighboring modes as in Figure~\ref{fig:MPS_tensor_intro}. 
In this work, we make use of the \textit{Time Evolving Density matrix with Orthogonal Polynomials Algorithm} (TEDOPA) procedure \cite{Tamascelli2020,Prior2010, chin_exact_2010}. This approach allows to transform the initial star conformation into a chain, via a change of basis (unitary transformation) based on orthogonal polynomials. In brief, given an environment characterized by a continuous spectral density $J(\omega)$, we express the orginal environmental modes in a new set of bosonic modes $\{c_n,c_n^{\dagger}\}$ as $a_k=\sum_{n=0}^\infty \sqrt{J(\omega_k)}\pi_n(\omega_k)c_n$, where $\pi_n(\omega)$ is a $n$th-order orthogonal polynomial w.r.t. to the ‘weight function' $J(\omega)$ \cite{chin_exact_2010}. Substituting this into the original Hamiltonian, and using the three-term recurrence properties of orthogonal polynomials, transforms $H_B$ into its chain form $H_C$
\begin{equation}
 H_C = \sum_{n=0}^\infty \epsilon_n c_n^{\dagger} c_n +t_n c_n^{\dagger}c_{n+1}+t_n c_n c_{n+1}^{\dagger},  
\end{equation}
where all microscopic information about the environment, i.e. $J(\omega)$ is now encoded in the chain parameters $\{\epsilon_n,t_n\}$. In certain cases, such as the widely used Ohmic, sub-Ohmic and super-Ohmic spectral densities, these chain parameters can even be determined analytically \cite{chin_exact_2010}, but efficient numerical procedures implemented in MPSDynamics.jl allow these parameters to be extracted for spectral densities with arbitrary frequency dependencies.   

\begin{figure}
    \centering
        \includegraphics[scale=1.0]{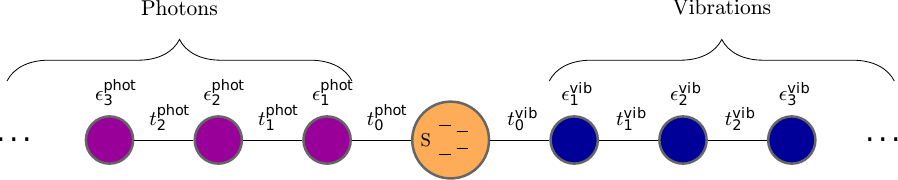}
    \caption{Chain conformation. The electronic system (orange) is interacting with two linearized environments : a chain of N photons (magenta) characterised by the parameters $\{\epsilon_n^\text{phot},t^\text{phot}_{n-1}\}$ with $n \in [1,N]$ and a chain of M vibrations (dark blue) characterised by $\{\epsilon_m^\text{vib},t^\text{vib}_{m-1}\}$ ($m \in [1,M]$). In this conformation, the system is interacting with only two modes and MPS techniques can be used.}
    \label{fig:4lvl_chain}
\end{figure}

It can further be shown that in the chain representation the system interacts with only the first mode of the chain-like environment and MPS techniques can then be easily applied. The MPS is then finally built as the system surrounded by a first chain of photons and a second chain of molecular vibrations as shown in Figure~\ref{fig:4lvl_chain}. This practice does not change the simulated wave-function but only the way to calculate dynamics. Transforming it into a chain allows to reach the full power of MPSs for a linear vibronic coupling. Moreover, the inverse mapping can always be performed, allowing us to use the chain results to extract observables and dynamics of the original bath degrees of freedom \cite{lacroix2024non,schroder2016simulating,riva2023thermal}. Indeed, we will make extensive use of this back-transformation throughout this article.

\subsection{Time-dependent Approach}

To simulate photo-chemistry process, it can be necessary to access ultrafast dynamics: initialising the wavepacket onto the excited state at the beginning of the simulation might wipe out rapid features. For this purpose, the time-dependent Hamiltonian $H_\text{drive} \left(t \right)$ is introduced within the MPS method in this work, representing a laser driving (either continuous or pulsed) that promotes the electronic population to the excited state. \\
        
In order to do so, the Hamiltonian is written as a Matrix Product Operator (MPO) and is modified at each time step to let the time-dependent part evolve. Practically, the time-dependent terms of the MPO are added to the time-independent MPO terms at each timestep in order to avoid the calculation of an entire new MPO. In this way, only a matrix sum is computed at each time step. Moreover, in cases where the energy gap between ground and excited states is large enough, the rotating wave approximation could easily be introduced to increase the simulation timestep and reduce the computational load.  




The wide possibilities of the generic four-level setup described by equations \eqref{eq:4lvl} to \eqref{eq:4lvl_end}, treated within the MPS framework and including the time-dependent feature, are illustrated on several model examples in the following paragraphs.

\subsection{Excited-State Dynamics and Dual-Fluorescence without Laser Driving}

\begin{figure}
    \centering
    \includegraphics[scale=1.0]{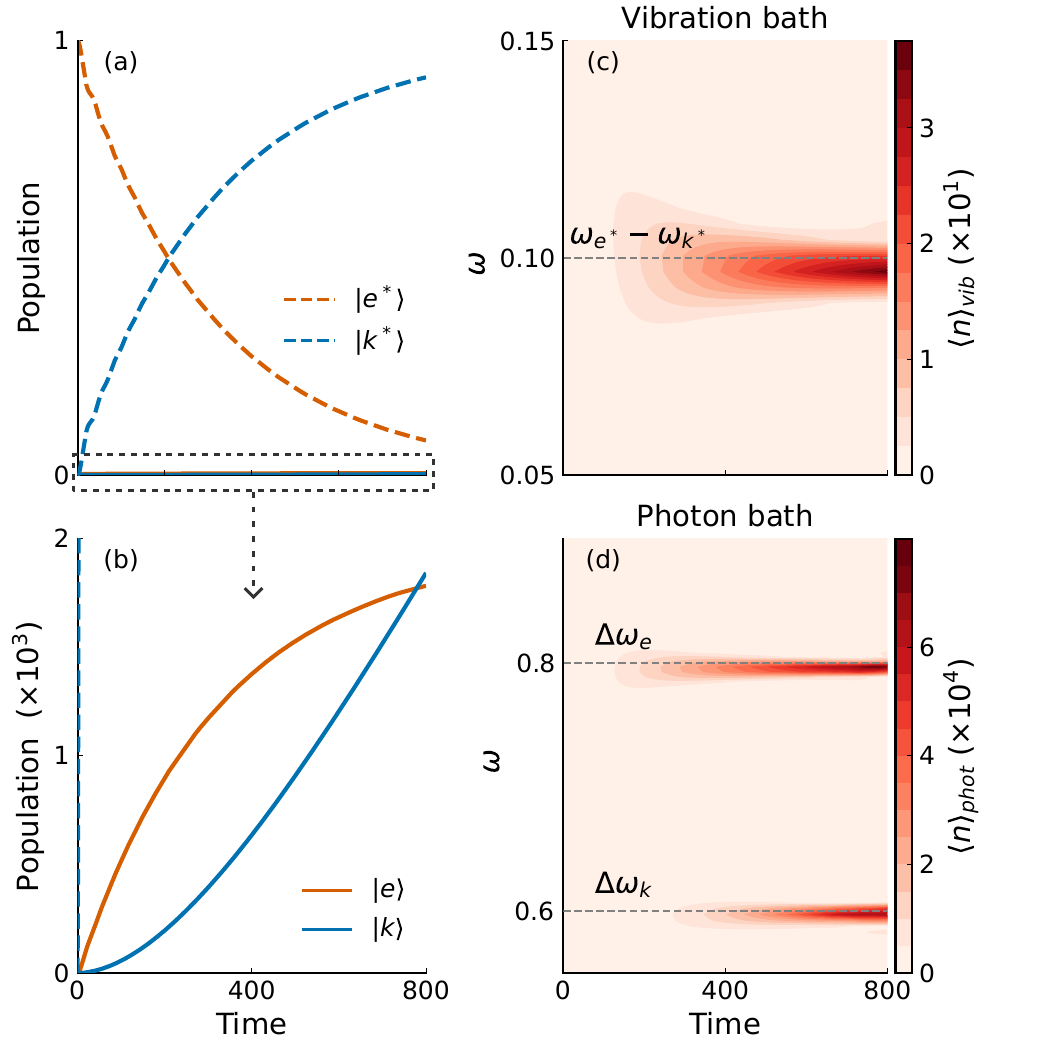}
    \caption{Dual-fluorescence example of a four-level system. (a) Dynamics of the electronic population. The dynamics starts in $\ket{e^*}$ and vibrations allow a transition towards $\ket{k^*}$. (b) Zoom in the dynamics of the electronic population. Photons allow to populate ground states. (c) Vibration average population resolved in time. An horizontal dashed line indicates the frequency difference $\omega_{e^*} - \omega_{k^*}$. (d) Photon average population resolved in time. Two different frequencies are detected, corresponding to the two electronic energy gaps. Time and frequencies are in arbitrary units.}
    \label{fig:dual_fluo_4lvl}
\end{figure}

First, we consider a simple four-level model where the initial population was prepared in the $\ket{e^*}$ state. The system here interacts with vibrations and photons and is governed by the Hamiltonian of eq.~\eqref{eq:4lvl} to \eqref{eq:4lvl_end}. In this first example, the Hamiltonian is time independent ($\epsilon(t) = 0$). All the parameters are reported in Table S1 of the Supporting Information. Beginning the dynamics with the $\ket{e^*}$ state populated, the electronic population is transferred from $\ket{e^*}$ to $\ket{k^*}$ as shown in Figure~\ref{fig:dual_fluo_4lvl} (a). Ground states $\ket{e}$ and $\ket{k}$ are also populated to a smaller extend via spontaneous photon emission (Figure~\ref{fig:dual_fluo_4lvl} (c)). Looking at the baths, vibration modes are populated in time mainly at one frequency (Figure~\ref{fig:dual_fluo_4lvl} (b)) while two frequencies are excited in the photon bath (Figure~\ref{fig:dual_fluo_4lvl} (d)). \\

This example illustrates the effect of environments on the electronic system. While there is no direct coupling between electronic states in $H_S$, the coupling to the vibration bath allows transitions from $\ket{e^*}$ to $\ket{k^*}$ as shown in Figure~\ref{fig:dual_fluo_4lvl} (a). Because of the energy transfer between the system and vibrations, vibration modes are excited at the frequency corresponding to the energy difference between $\ket{e^*}$ and $\ket{k^*}$ (Figure~\ref{fig:dual_fluo_4lvl} (c)). 
Coupling to the photon bath also triggers vertical transitions, thereby populating the electronic ground states (Figure~\ref{fig:dual_fluo_4lvl} (b)). Two emission lines can be observed, increasing in intensity with time (Figure~\ref{fig:dual_fluo_4lvl} (d)). These peaks are characteristic of dual fluorescence, i.e. simultaneous emission from the enol (at frequency $\Delta \omega_e$) and from the keto forms (at frequency $\Delta \omega_k$). The keto emission is delayed in comparison with the enol one because of the different initial population of the excited states. Although ESIPT happens in a sub-picosecond timescale, this time-resolved photo-emission spectrum can, in principle, be probed experimentally \cite{lee_active_2013,kim_coherent_2009}. This example shows that, by making use of the MPS wave-function method, information can be recovered from environments. All of these observables together enable the energy transfers to be tracked, both in the system and in the environments. \\

\subsection{Excited-State Dynamics under Weak and Strong Laser Driving}
\label{subsec:4lvl-driving}

We now consider a dynamics initiated in the enol ground state $\ket{e}$ and introduce the effect of the driving field via the time-dependent term in order to populate the excited state (eq.~\ref{eq:4lvl_hst}). The system and bath characteristics are similar to the previous example, while the driving is continuous and relatively weak ($\epsilon = 2 \pi/800 ~$arb. units). The field frequency is set up at $\omega_\text{drive} = \Delta \omega_{e}$ in order to populate the enol excited state and triggers the first step of the ESIPT process. Since the keto energy gap is off resonance, the continuous drive has a negligible effect on the keto states. For parameters, see Table S2 of the Supporting Information. \\

\begin{figure}
    \centering
    \includegraphics[scale=1.0]{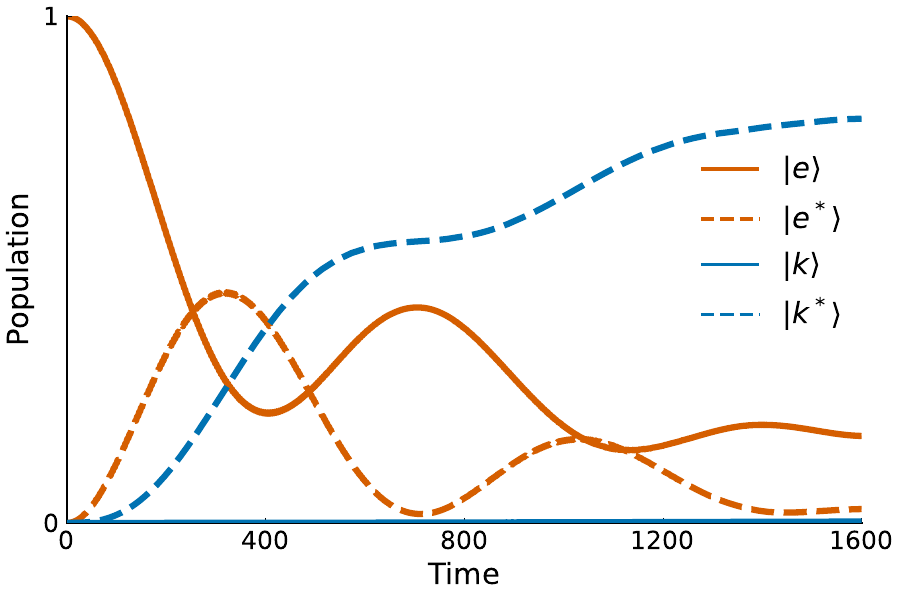}
    \caption{Weakly-driven four-level system. The electronic dynamics starts in $\ket{e}$ and $\ket{e^*}$ is populated with the pulse. Transfer in $\ket{k^*}$ happens due to the coupling between $\ket{e^*}$ and $\ket{k^*}$ with vibrations. Time is in arbitrary units.}
    \label{fig:weakdriving_4lvl}
\end{figure}

The electronic population dynamics is presented in Figure~\ref{fig:weakdriving_4lvl}. The $\ket{e}$ and $\ket{e^*}$ populations show damped oscillations in time, with a characteristic period close to $800~$arb. units, corresponding to the Rabi period of the drive. In the long term, the total enol population goes to zero, while the keto population $\ket{k^*}$ increases towards 1. Vibrations allow transfers between $\ket{e^*}$ and $\ket{k^*}$ so that $\ket{k^*}$ becomes populated as soon as $\ket{e^*}$ is populated by the drive. In the long term, the population is entirely transferred from enol to keto via the excited states. Environment spectra for this example are not shown but they closely resemble that of Figure~\ref{fig:dual_fluo_4lvl} .\\

\begin{figure}
    \centering
    \includegraphics[scale=1.0]{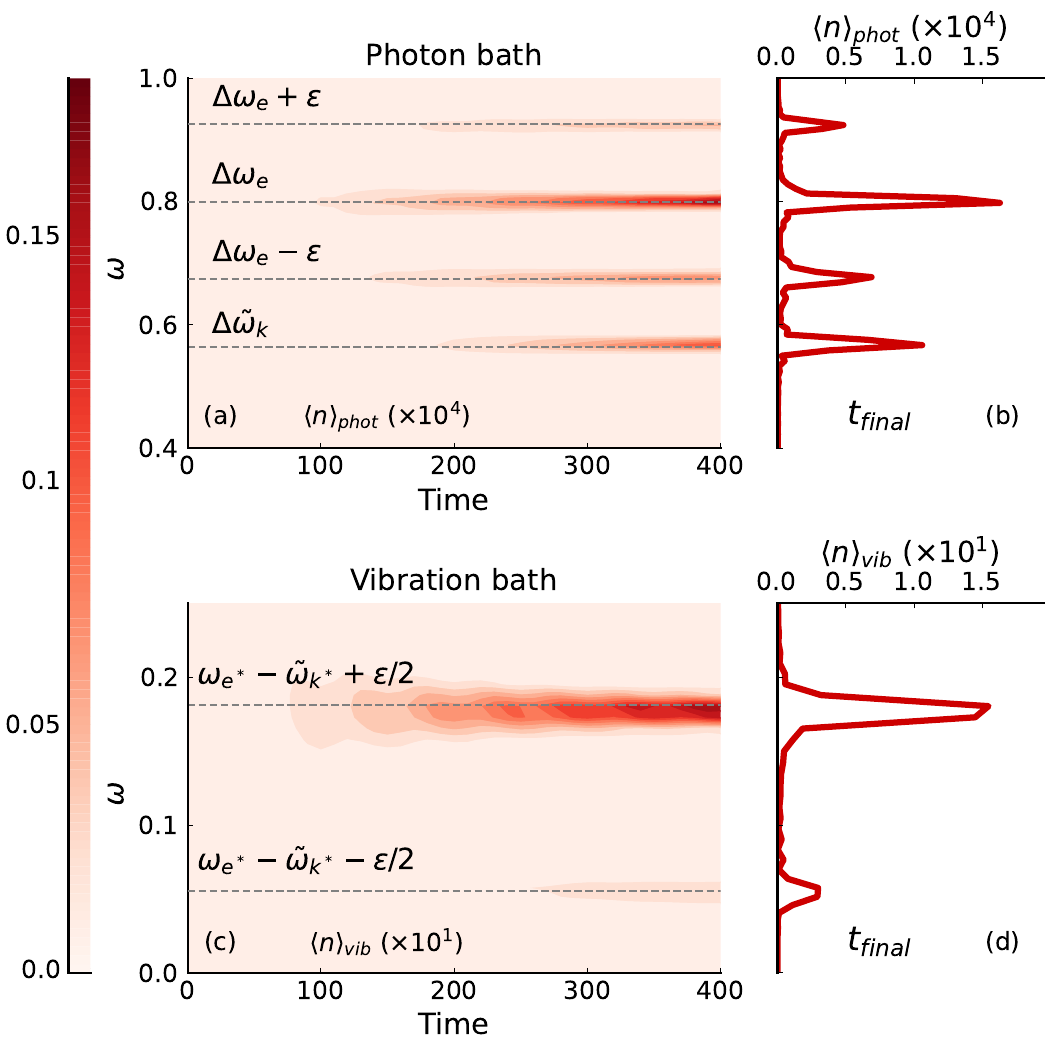}
    \caption{Strongly-driven four-level system. (a) Photon average population resolved in time. A Mollow triplet appears alongside the keto transition. (b) Photon average population at $t_\text{final}=400~$arb. units. (c) Vibration modes average population resolved in time. Due to the strong driving, two transitions are now possible between $\ket{e^*}$ and $\ket{k^*}$. (d) Vibration modes average population at $t_\text{final}=400~$arb. units. Time and frequencies are in arbitrary units.}
    \label{fig:mollowtriplet_4lvl}
\end{figure}

As a last example, we consider the same system under a strong driving $\epsilon = 2 \pi/50~$arb. units (all parameters are reported in Table S3 of the Supporting Information). This value corresponds to strong light-matter coupling, that could be reached by the means of an intense laser excitation or in a resonant optical cavity. 
Figure~\ref{fig:mollowtriplet_4lvl} shows the excitation spectrum for the photon environment (Figure~\ref{fig:mollowtriplet_4lvl} (a),(b)) and for the vibration environment (Figure~\ref{fig:mollowtriplet_4lvl} (c),(d)). Four main photo-emission lines can be identified in the photon bath results, and two different molecular vibration frequencies are excited by the system. The corresponding electronic population dynamics is reported in the Supporting Information (Figure~S1).\\

\begin{figure}
    \centering
    \includegraphics[scale=1.2]{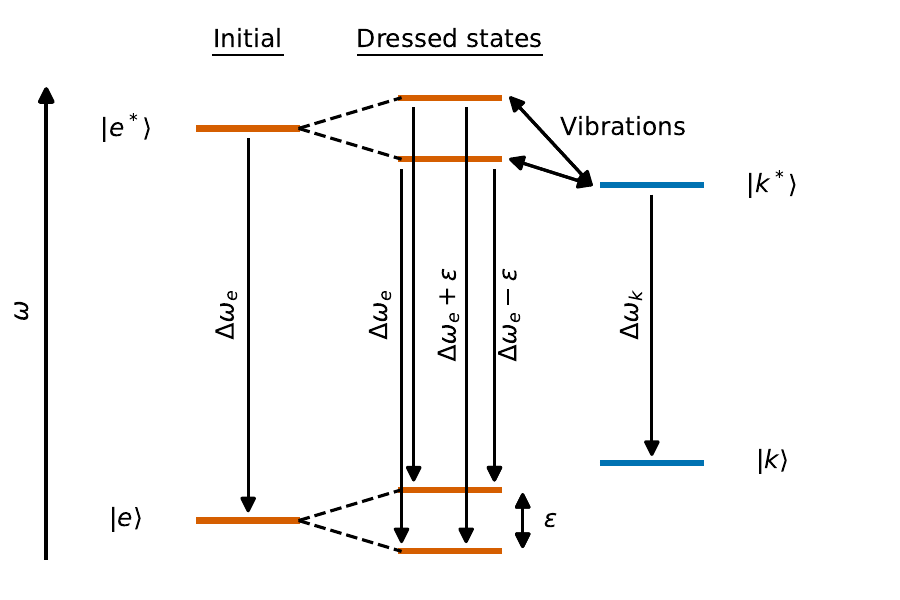}
    \caption{Four-level system strongly driven on resonance with $\Delta \omega_e$. It creates enol dressed states with a new $\epsilon$ gap and three different optical transitions, known as a Mollow triplet. Two different transitions between the dressed states $\ket{e^*}$ and the state $\ket{k^*}$ are now possible.}
    \label{fig:mollowtriplet_scheme}
\end{figure}

The formation of these peaks is characteristic of the strong coupling regime and can be explained analytically: the strong coupling between the $\ket{e}\rightarrow \ket{e^*}$ transition and the driving field gives rise to enol dressed states. As a result, the $\ket{e}\rightarrow \ket{e^*}$ emission line splits in three distinct peaks, corresponding to transitions between the different dressed states (Fig.~\ref{fig:mollowtriplet_scheme}). This characteristic signature of strong light-matter coupling is known as Mollow triplet \cite{mollow_power_1969}. The central peak corresponds to the two degenerate transitions that remain at the initial frequency $\Delta \omega_e$, while two other transitions are found at frequencies $\Delta \omega_e + \epsilon$ and $\Delta \omega_e - \epsilon$. Furthermore, the state $\ket{k^*}$ now interacts with the two $\ket{e^*}$ dressed states, which also explains the splitting of the vibration spectrum peak. Finally, though the coupling between the driving field and the keto states is not on resonance, it still has a measurable impact on these states and results in a slight shift of the keto energy gap $\Delta \tilde{\omega}_k$. All the shifts arising from the strong coupling and the emergence of dressed states can be anticipated on the basis of a simple analytical model that can be found in Supplementary Information.

In Figure~\ref{fig:mollowtriplet_4lvl}, the lowest frequency photon signal corresponds to the dressed transition $\ket{k^*} \rightarrow \ket{k}$ (Figure~\ref{fig:mollowtriplet_4lvl} (a), (b)) while the three other peaks constitute the Mollow triplets. The highest central peak corresponds to the two degenerate enol transitions at $\Delta \omega_e$ and the two satellite peaks are at $\Delta \omega_e + \epsilon$ and $\Delta \omega_e - \epsilon$. High frequency transitions are favoured by the vibrational spectral density, which explains the slight difference in intensity between the Mollow triplet satellites: the upper enol dressed state being more strongly coupled to $\ket{k^*}$, it depopulates more quickly and contributes less to the photo-emission spectrum. We also note that the keto emission is differed in time from the enol emission, as a result of the delay induced by the $\ket{e^*} \rightarrow \ket{k^*}$ proton transfer step. \\

 The two peaks in the vibration spectrum (Figure~\ref{fig:mollowtriplet_4lvl} (c), (d)) correspond to transitions from the two dressed states of $\ket{e^*}$ to $\ket{k^*}$. Two excited-state transitions are now possible at $\omega_{e^*} - \tilde{\omega}_{k^*} + \epsilon/2$ and at $\omega_{e^*} - \tilde{\omega}_{k^*} - \epsilon/2$ (Figure~\ref{fig:mollowtriplet_scheme}). Finally, strong coupling with the driving field gives rise to Rabi oscillations in time in both the vibration and photon environments. These oscillations are shown in more detail in Supplementary Information (Figure~S2). 

\newpage

\section{From Discrete States to Non-Adiabatic Dynamics on Continuous Energy Landscapes}
\label{sec:RC}

The ESIPT process involves not only an electronic transition between the states $\ket{e^*} $ and $ \ket{k^*}$, but also a proton transfer. As the two phenomena are highly correlated, an explicit description of both mechanisms is crucial \cite{harshan_dependence_2015}. In contrast, in the previous 4-state model, the proton transfer is only described implicitly in the $\ket{e^*} \rightarrow \ket{k^*}$ transition, mediated by the coupling with the vibration environment. The MPS formalism is generally not used to describe continuous reaction coordinates such as the position of the transferring proton. In this section, we develop a model that allows the consistent description, within the MPS Ansatz, of the proton wavefunction along its path from the donor to the acceptor site. The associated continuous reaction coordinate (RC) is described by means of an additional vibration mode, with appropriate couplings to the electronic states in order to generate the double-well energy profile typical of proton transfer processes. 


\subsection{Hamiltonian}
 
To set up the RC feature in the MPS method, additional terms are included in the Hamiltonian. In this section, we focus on the electronically excited states to simplify the description, the ground states $\ket{e}$ and $\ket{k}$ can be treated in exactly the same way. 
A coupling term $\Delta$ is introduced between the electronic states $\ket{e^*}$ and $\ket{k^*}$, while the RC mode is described by the oscillator annihilation and creation operators $d, \ d^{\dagger}$ with a characteristic frequency $\omega_\text{RC}$~:
\begin{subequations}
\renewcommand{\theequation}{\theparentequation.\arabic{equation}}
\begin{align} \label{eq:RC_wo_bath}
    &H = H_{\text{S}} + H_{\text{RC}} + H_{\text{int}}^{\text{RC-S}} \\ \label{eq:TLS_RC}
    &H_{\text{S}} = \omega_{e^*}\ket{e^*}\bra{e^*} + \omega_{k^*}\ket{k^*}\bra{k^*} + \Delta \left( \ket{e^*}\bra{k^*} + \ket{k^*}\bra{e^*} \right) \\ 
    &H_{\text{RC}} = \omega_{\text{RC}} \left(d^{\dagger}d + \frac{1}{2} \right)\\
    &H_{\text{int}}^{\text{RC-S}} = g_{e^*} \ket{e^*}\bra{e^*}\left( d + d^{\dagger}\right)+ g_{k^*} \ket{k^*}\bra{k^*}\left( d + d^{\dagger}\right)
\end{align}
\end{subequations}
The coupling coefficients $g_{e^*}$ and $g_{k^*}$ between the electronic and RC parts of the Hamiltonian shift the RC equilibrium position in a different way for the enol and keto forms, therefore giving rise to the characteristic double-well potential of proton transfer. \\

\begin{figure}
         \centering
         \includegraphics[width=1\textwidth]{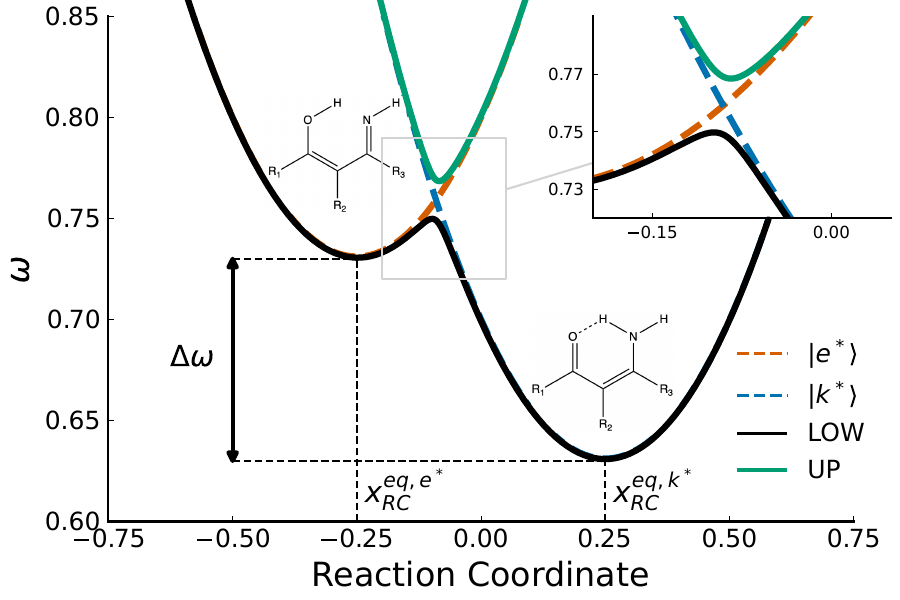}
        \caption{Example of potential energy curves parameterised by the Hamiltonian of eq~\ref{eq:RC_wo_bath} in the adiabatic basis. The Hamiltonian is diagonalized leading to two distinct adiabatic surfaces, that are designed as LOW and UP surfaces. Diabatic surfaces are also represented underneath as dashed lines. We define $\Delta \omega = \omega_{e^*} - \omega_{k^*} + \frac{1}{2}m_\text{RC} \omega_\text{RC}^2 \left( (x_\text{RC}^{\text{eq,}k^*})^2 - (x_\text{RC}^{\text{eq,}e^*})^2 \right)$. The Reaction Coordinate and $\omega$ are in arbitrary units.}
\label{fig:DiabAdiab}
\end{figure}

\noindent \textbf{Continuous description in space.} A more intuitive representation of the system is indeed obtained by re-expressing the Hamiltonian on the basis of the RC spatial coordinate. We first recall that the position operator is defined from the annihilation and creation operators as $x=\sqrt{\frac{1}{2m_\text{RC} \omega_{\text{RC}}}}\left(d + d^{\dagger} \right)$. The Hamiltonian of eq~\ref{eq:RC_wo_bath} can then be rewritten as:
\begin{equation} \label{eq:RC_wo_bath_space}
    H = H^{e^*}(x)\ket{e^*}\bra{e^*} +H^{k^*}(x)\ket{k^*}\bra{k^*} + \Delta \left( \ket{e^*}\bra{k^*} + \ket{k^*}\bra{e^*} \right) 
\end{equation}
\noindent where $H^{e^*}(x) = \frac{1}{2} m_\text{RC} \omega_\text{RC}^2 \left( x - x_\text{RC}^{\text{eq,}e^*}\right)^2 + \omega^{\text{eq},e^*}$ 
and similarly for $H^{k^*}(x)$. The equilibrium positions $x_\text{RC}^{\text{eq,}e^*}$ and $x_\text{RC}^{\text{eq,}k^*}$ of the excited enol and keto states are directly related to the interaction coefficients via : $g_i = - \omega_\text{RC}  \, x_\text{RC}^{\text{eq,}i} \sqrt{\frac{m_{\text{RC}}\omega_\text{RC}}{2}} $ and the energies at the two equilibrium positions follow, $\omega^{\text{eq},i}= \omega_i-\frac{1}{2} m_\text{RC} \omega_\text{RC}^2 \left(x_\text{RC}^{\text{eq,i}}\right)^2$.   
In this representation, the system can be seen as consisting in two displaced harmonic oscillators for $\ket{e^*}$ and $\ket{k^*}$, with different equilibrium positions and energies and coupled via the tunnelling term $\Delta$. The different parameters enable a rich variety of configurations for the corresponding potential surface, an example is represented by the dashed lines in Figure~\ref{fig:DiabAdiab}. 


\noindent \textbf{Diabatic and adiabatic basis.} Two different bases are frequently used to represent this type of systems at a fixed RC position $x$. The initial Hamiltonian of eq~\ref{eq:RC_wo_bath_space} is written in the so-called diabatic electronic basis $\left(\ket{e^*}, \ket{k^*} \right)$ and represented by the following matrix:
\begin{equation}
    H^{\text{Diab}}(x)=\begin{pmatrix}
    H^{e^*}(x)  & \Delta\\ 
    \Delta & H^{k^*}(x)  \\
\end{pmatrix}
\end{equation}
where the anti-diagonal terms are the coupling between the electronic states. The adiabatic basis is then obtained by diagonalizing $H^{\text{Diab}}(x)$ at fixed position $x$:
\begin{equation}
H^{\text{Adiab}}(x)=\begin{pmatrix}
    H^\text{LOW}(x)  & 0\\
    0 & H^\text{UP}(x)  \\
\end{pmatrix}
\end{equation}
The adiabatic energies $H^\text{LOW}(x)$ and $ H^\text{UP}(x)$ are represented in solid lines in Figure.~\ref{fig:DiabAdiab}. The corresponding eigenvector $\ket{\chi^\text{LOW}(x)}$ and $\ket{\chi^\text{UP}(x)}$ follow from the diagonalization of the matrix $H^{\text{Diab}}(x)$. 

The lower potential surface $H^\text{LOW}(x)$ has a characteristic double-well shape. In the following, the RC is treated at a quantum level using the MPS representation, its dynamics therefore describes the evolution of the proton wavepacket, propagating onto the double-well potential energy surface after being photo-excited. 

\subsection{The Adiabatic Projector}
\label{subsec:adiabproj}
The inclusion of the continuous RC variable leads to an extended system described in terms of electronic state and spatial coordinate, as shown in Figure~\ref{fig:DiabAdiab}. The MPS method is naturally expressed in the diabatic basis, whereas most photo-chemistry problems are generally described in the adiabatic basis \cite{Frster1970,Turro1979}. For this reason, we developed an adiabatic projector (expressed in the MPO representation) that enables easy transformations from one basis to the other. \\

\noindent \textbf{Construction of the adiabatic projector.} The purpose of the adiabatic projector is to split the system's wavepacket into two parts : the one evolving onto the upper adiabatic surface and the other one onto the lower surface. The projector is built as an MPO, which allows for an efficient numerical treatment. 


In order to build the projector, we first consider the normalised eigenstates of the undisplaced RC harmonic oscillator $H_{\text{RC}}$. These are the Fock states $\ket{n}$ over which the MPS dynamics is build, with the associated eigenfunctions $\phi_n(x)$ given by the usual Hermite polynomials. A large number of Fock states should be considered in order to obtain a well-resolved spatial description of the RC. The second ingredient to the adiabatic projector is the eigenstates of the diagonalization from the diabatic to the adiabatic electronic basis at fixed $x$ value: $\ket{\chi^\text{LOW}(x)}$ and $\ket{\chi^\text{UP}(x)}$. A generic system wavefunction can then be decomposed as: 
\begin{subequations}
\renewcommand{\theequation}{\theparentequation.\arabic{equation}}
\begin{align}
    \ket{\psi^{\text{Adiab}}(x,t)} &= \ket{\psi^{\text{LOW}}(x,t)} + \ket{\psi^{\text{UP}}(x,t)} \\
    \ket{\psi^{\text{Adiab}}(x,t)} &= \ket{\chi^\text{LOW}(x)}\phi_{\text{LOW}}(x,t) + \ket{\chi^\text{UP}(x)}\phi_{\text{UP}}(x,t)
\end{align}
\end{subequations}
with $\phi_{\text{LOW}}(x,t)$ and $\phi_{\text{UP}}(x,t)$ the time-dependent adiabatic wavefunctions and $\ket{\chi^\text{LOW/UP}(x)} = \chi_{e^*}^\text{LOW/UP}(x) \ket{e^*} + \chi_{k^*}^\text{LOW/UP}(x) \ket{k^*}$ the local electronic eigenstates. 
The projector $\hat{P}_{\text{AD}}^{\text{LOW}}$ on the LOW adiabatic surface can then be written as (a similar form holds for the UP projector): 
\begin{subequations}
\renewcommand{\theequation}{\theparentequation.\arabic{equation}}
\begin{align}
    \hat{P}_{\text{AD}}^{\text{LOW}} &= \int \mathrm{d}x \ket{x}\bra{x} \otimes  \ket{\chi^\text{LOW}(x)}\bra{\chi^\text{LOW}(x)}\\
    &= \sum_{n,m} \int \mathrm{d}x \, \phi_n(x) \phi_m^*(x)\ket{n}\bra{m} \otimes  \left( \chi_{e^*}^\text{LOW}(x)\ket{e^*} + \chi_{k^*}^\text{LOW}(x)\ket{k^*} \right) \otimes \\
     &\left(\chi_{e^*}^{\text{LOW}*}(x)\bra{e^*} + \chi_{k^*}^{\text{LOW}*}(x)\bra{k^*} \right)
\end{align}
\noindent In the diabatic basis over which the MPS dynamics is run, the projector can thus be written as:
\begin{equation}\label{Proj}
    \hat{P}_{\text{AD}}^{\text{LOW}} = \sum_{n,m,i,j} \ket{n}\bra{m} \otimes  \ket{i}\bra{j} \Delta_{nm}^{ij}
\end{equation}
\end{subequations}
with
\begin{equation}
    \Delta_{nm}^{ij} = \int \mathrm{d}x \,  \phi_n(x) \phi_m^*(x) \chi_{i}^\text{LOW}(x) \chi_{j}^{\text{LOW}*}(x) 
\end{equation}
where $i$ and $j$ are either $e^*$ or $k^*$ while $n$ and $m$ represent the different Fock states of the RC harmonic oscillator. \\

\noindent \textbf{MPO formulation of the adiabatic projector.} From the MPO perspective, the projector can be seen as acting only on two sites: that of the electronic state $\bra{e^*}$, $\bra{k^*}$ and that of the RC oscillator. The corresponding tensor therefore combines the electronic tensor with the RC mode tensor. For instance, if we focus on the projection onto the LOW surface with $\hat{P}_{\text{AD}}^{\text{LOW}}$, and define a discrete spatial grid $x=(x_1,x_2,\ldots,x_N)$ and the electronic transition matrix $M^{\text{LOW}}$ for each discretized space point $x_i$ ($i \in [1,N]$):
   \begin{equation}   
    M^{\text{LOW}}(x_i)=\begin{pmatrix}
    \chi^{\text{LOW}}_{e^*}(x_i) \chi^{*,\text{LOW}}_{e^*}(x_i) & \chi^{\text{LOW}}_{e^*}(x_i) \chi^{*,\text{LOW}}_{k^*}(x_i) \\
    \chi^{\text{LOW}}_{k^*}(x_i) \chi^{*,\text{LOW}}_{e^*}(x_i) &  \chi^{\text{LOW}}_{k^*}(x_i) \chi^{*,\text{LOW}}_{k^*}(x_i)
\end{pmatrix}
\end{equation}
For a number $M$ of Fock states, we also define the $P(x_i)$ matrix from the products of eigenstates at fixed $x_i$: 
\begin{equation}
    P(x_i) = \begin{pmatrix}
        \phi_1(x_i)\phi^*_1(x_i) & \ldots & \phi_M(x_i)\phi^*_1(x_i) \\
        \vdots & \, & \vdots \\
        \phi_1(x_i)\phi^*_M(x_i) & \ldots & \phi_M(x_i)\phi^*_M(x_i) \\
    \end{pmatrix}
\end{equation}

\noindent Now, the full MPO projector can be written as:
 \begin{equation}\label{eq:MPO_proj}
     \hat{P}_{\text{AD}}^{\text{LOW}} = \underbrace{\mathbb{1} \otimes \ldots \otimes \mathbb{1} }_{\text{Photons}} \otimes \underbrace{\begin{pmatrix}
         M^{\text{LOW}}(x_1) & M^{\text{LOW}}(x_2) & \ldots & M^{\text{LOW}}(x_N)
     \end{pmatrix}}_{\text{S}} \otimes \underbrace{\begin{pmatrix}
         P(x_1) \\
         P(x_2) \\
         \vdots \\
         P(x_N)
     \end{pmatrix}}_{\text{RC mode}} \otimes \underbrace{\mathbb{1} \otimes \ldots \otimes \mathbb{1}}_{\text{Vibrations}}
 \end{equation}


Expressed in this way, the projector combines the adiabatic electronic states obtained for each discrete position $x_i$.  
The dimension of the associated MPO should therefore scale linearly with the number of grid points $x_i$ as it is clear from eq~\ref{eq:MPO_proj}. This should be numerically cumbersome, as a fine grid might be needed for an accurate description of the RC wavepacket. 
This manoeuvre would dramatically increase the bond dimension of the resulting wave-function, as the product of a MPS with an MPO is an MPS whose bond dimension is the \emph{product} of the original MPS' and the MPO's bond dimensions \cite{schollwock2011density}. One way to circumvent this unnecessary complexity is to reduce the dimension of the projector, by applying a Singular Value Decomposition (SVD), which reduces the projector operator to its necessary elements, while the result of its application to a wavefunction is not affected. 
More precisely, the bond dimension between the electronic MPO tensor and the RC mode MPO tensor is the number of discrete grid points $x_i$. This number can be large when Fock states of high index have to be taken into account. 
In the following examples, corresponding to typical ESIPT four-level systems, the number of $x$ points varies between $1300$ and $2800$. After SVD, the bond dimension of the projector is reduced to $5$ for both the symmetric and non-symmetric double-well cases.

Once expressed in the MPO formalism and reduced using the SVD, the adiabatic projectors $\hat{P}_{\text{AD}}^{\text{LOW}}$ and $\hat{P}_{\text{AD}}^{\text{UP}}$ provide a useful and efficient tool to finely analyze the dynamics of the system, as illustrated by the examples of application in the following.


\subsection{Coupling between Vibrations and the Reaction Coordinate}

The MPS approach allows not only to simulate the quantum dynamics of the isolated \{Electrons + Reaction Coordinate\} system but also to efficiently describe its coupling with the vibrational and photonic environments. In particular, the vibrational environment dissipates energy away from the system, acting as a damping on the continuous RC coordinate that causes the wavepacket to relax to the lower energy states. 


In practice, the RC oscillator is coupled to a set of harmonic oscillators of mass $m$. The coupling Hamiltonian takes the characteristic form: 
\begin{subequations}
\renewcommand{\theequation}{\theparentequation.\arabic{equation}}
\begin{align}
       H_{\text{vib}} + H_{\text{int}}^{\text{vib-RC}} &= \left( \sum_k \frac{p_k^2}{2m} + \sum_k \frac{1}{2} m \omega_k^2 \left(q_k - \alpha_k \, x_{\text{RC}} \right)^2\right)
\end{align}
with $q_k = \sqrt{\frac{1}{2m\omega_k}}\left(b_k + b_k^{\dagger} \right)$ the position operator of the $k^\text{th}$ vibration mode of frequency $\omega_k$ and $p_k = i \sqrt{\frac{m\omega_k}{2}}\left(b_k^{\dagger} - b_k \right)$ its momentum. The interaction coefficients $\alpha_k$ are potentially different for each mode which enables representing arbitrary spectral densities for the vibration bath. To express this Hamiltonian in the MPS frame, it needs to be converted in second quantization:
\begin{align}
        H_{\text{vib}} &= \sum_k \omega_k b_k^{\dagger} b_k \\
       H_{\text{int}}^{\text{vib-RC}} &= \sum_k \frac{1}{2} m \omega_k^2 \left(-2 \, \alpha_k x_{\text{RC}} q_k + \alpha_k^2 x_{\text{RC}}^2\right) \\
          &= \sum_k \frac{\alpha_k^2 m \omega_k^2}{4m_\text{RC}\omega_\text{RC}} \left(d + d^{\dagger} \right)^2 - \left(d + d^{\dagger} \right) \sum_k \frac{\alpha_k \,  \omega_k \sqrt{m \, \omega_k}}{2\sqrt{m_\text{RC} \, \, \omega_\text{RC} }}  \left(b_k + b_k^{\dagger} \right) \label{eq:RC_vibcoupl} \\
             &= \lambda_{\text{reorg}} \left(d + d^{\dagger} \right)^2 - \left(d + d^{\dagger} \right) \sum_k z_k \left(b_k + b_k^{\dagger} \right)  \label{eq:RC_vibcoupl2}
\end{align}
\end{subequations}
The final interaction Hamiltonian takes the form a linear coupling between the RC mode and an environment of vibrations, weighted by the coefficients $z_k = \frac{\alpha_k \,  \omega_k \sqrt{m \, \omega_k}}{2\sqrt{m_\text{RC} \, \, \omega_\text{RC} }}$ corresponding to a spectral density $J\left(\omega\right)=\sum_k z_k^2 \delta\left(\omega_k - \omega \right)$. A supplementary term arises from the development, designated as the reorganisation energy $\lambda_{\text{reorg}} =   \int \frac{J(\omega)}{\omega}\text{d}\omega $. This term compensates the modification of the double-well potential shape that would be induced in its absence by the vibrational environment. 

\begin{figure} 
    \centering
     \includegraphics[width=1.0\textwidth]{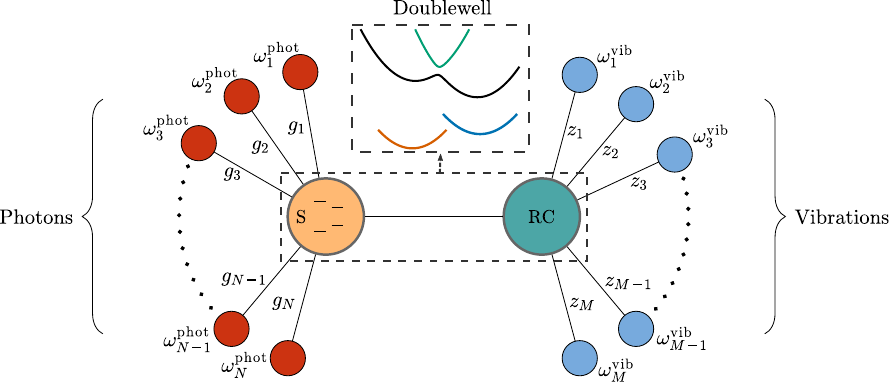}
    \caption{\{Electrons + Reaction Coordinate\} star conformation. The electronic system (orange) is interacting with $N$ photon modes (red) characterised by the spectral density $J_\text{phot}\left(\omega\right)=\sum_{k=1}^N g_k^2 \delta\left(\omega_k - \omega \right)$. The system is also coupled to the RC mode, forming the double-well. The RC mode is itself interacting with $M$ vibration modes (blue) characterised by the spectral density $J_\text{vib}\left(\omega\right)=\sum_{k=1}^M z_k^2 \delta\left(\omega_k - \omega \right)$.}
    \label{fig:MPS_RC}
\end{figure}

This type of coupling leads to the star conformation of Figure~\ref{fig:MPS_RC}. The electronic system is coupled to the RC mode which gives rise to a double-well potential energy surface. Environments include photons interacting with the electronic component of the system and molecular vibrations, coupled to the RC mode via equation~\eqref{eq:RC_vibcoupl2}. This picture also holds a physical meaning. The RC mode can be interpreted as one particular mode of the molecular vibrations that is strongly coupled to the electronic states. Therefore, this particular mode is treated seperately from the other skeleton vibrations of the vibration environment in order to gather more information on its quantum state.

\subsection{Reduced Density Matrix}

To take full advantage of continuous description of the reaction coordinate, one can compute the corresponding reduced density matrix. It is obtained by tracing out the two environments and projecting on one or the other of the adiabatic electronic states in order to obtain a description of the quantum wavepacket corresponding specifically to the RC mode. It allows a rich and visual study of the dynamics, going beyond mean values of observables. The full quantum state of the RC mode can be captured, including the spatial delocalisation of the transferring proton and the associated coherences (via the diagonal and off-diagonal elements of the reduced density matrix, respectively). 



\subsection{Dissipative Tunnelling in a Symmetric Double-Well Potential}

\begin{figure}
   \includegraphics[width=1.0\textwidth]{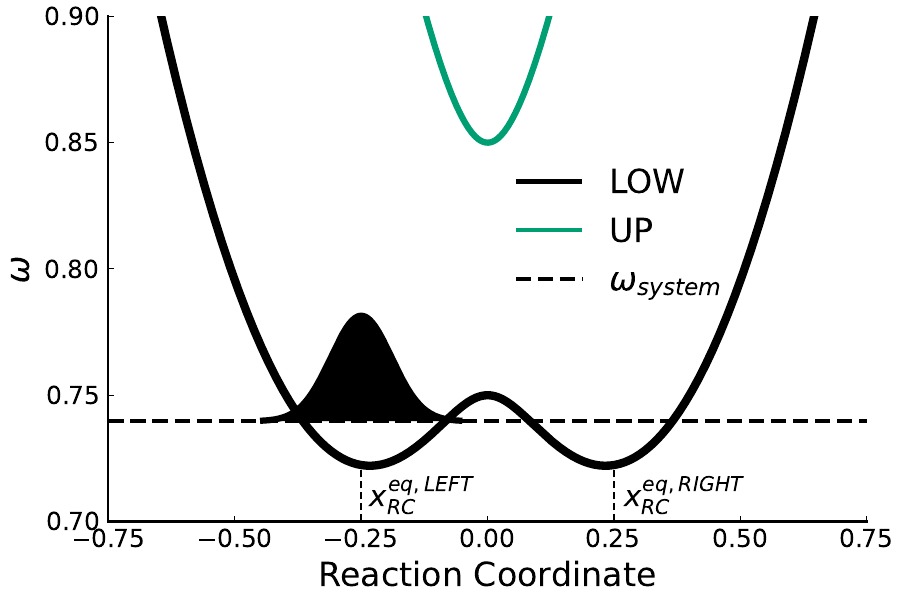}
\caption{Wavepacket (schematised in black) initially localised onto the LOW left well. The chosen Hamiltonian parameters design the adiabatic LOW and UP surface as well as $x_\text{RC}^\text{eq, LEFT}$ and $x_\text{RC}^\text{eq, RIGHT}$. The initial energy of the isolated system is here below the barrier. RC value and $\omega$ are in arbitrary units.}
\label{fig:tunneling_ini}
\end{figure}

To illustrate the different features presented in this section, a first elementary example describes a wavepacket that lies only onto the excited states. The wavepacket is initialised in the left well of a symmetric excited adiabatic LOW surface (Figure~\ref{fig:tunneling_ini}), whereas the high energy of the excited UP surface prevents any non-adiabatic transitions in the dynamics. For the sake of the example we call the diabatic parameters $x_\text{RC}^{\text{eq,}e^*}  = x_\text{RC}^\text{eq,LEFT}$ and $x_\text{RC}^{\text{eq,}k^*}  = x_\text{RC}^\text{eq,RIGHT}$. Parameters are reported in Table S4 and Table S5 of the Supporting Information.

\begin{figure}
   \includegraphics[width=1.0\textwidth]{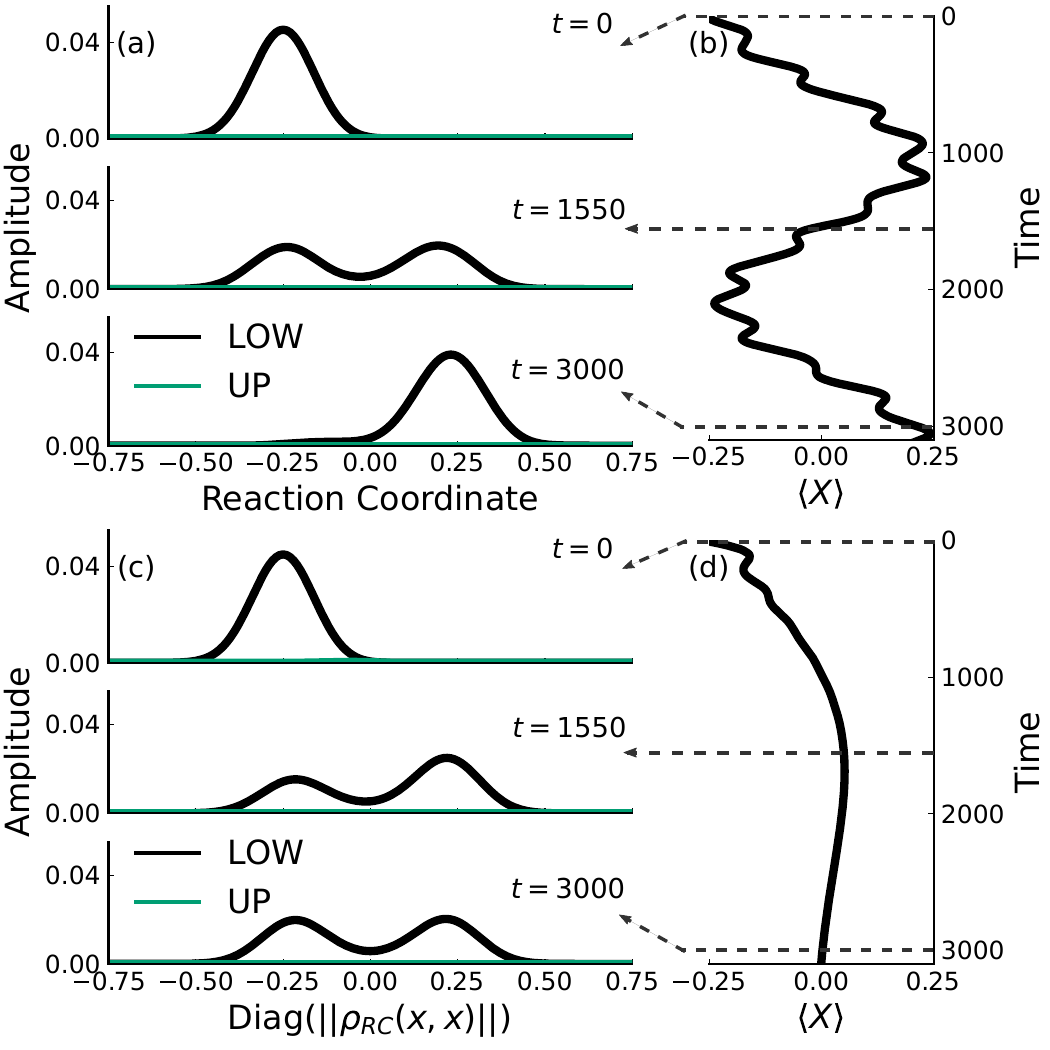}
\caption{Dynamics of a system initialised in the left well.(a) Time snapshots of the isolated system wavepacket. The LOW and UP wavepacket parts are obtained with the post-analysis adiabatic projector. The tunnelling effect allows passage between the left and the right well.(b) Mean value of the $X$ operator for the isolated system. Oscillations between minima of the left and right wells are shown.(c) Time snapshots of the adiabatic diagonal elements $||\rho_\text{RC}^\text{LOW/UP}(x,x)||$. Vibrations stabilise the system towards the ground state.(d) Mean value of the $X$ operator for the damped system. Vibrations make it converge towards $\langle X \rangle = 0~$arb. units.}
\label{fig:tunneling_dynamics}
\end{figure}

Figure~\ref{fig:tunneling_dynamics} shows the dynamics of the isolated system ((a),(b)) and of the system damped by vibrations ((c),(d)). Concerning the isolated system, Figure~\ref{fig:tunneling_dynamics} (a) shows time snapshots of the wavepacket amplitude along the Reaction Coordinate. We can see that the wavepacket is travelling back and forth between the left well and the right well. The mean value of the displacement operator $X$ is also represented in Figure~\ref{fig:tunneling_dynamics} (b) showing oscillations between $\langle X \rangle = x_\text{RC}^\text{eq,LEFT} = -0.25~$arb. units and $\langle X \rangle = x_\text{RC}^\text{eq,RIGHT} = 0.25~$arb. units with smaller oscillations. Once vibrations are introduced, wavepacket time snapshots (Figure~\ref{fig:tunneling_dynamics} (c)) show a stabilisation with the wavepacket equally split between the two wells. The $\langle X \rangle $ dynamics (Figure~\ref{fig:tunneling_dynamics} (d)) is also modified, converging towards $\langle X \rangle = 0~$arb units. \\


Regarding the isolated case, although the initial state has an energy lower than the potential barrier, the quantum description of the RC mode enables tunnelling to occur. The adiabatic projector used as a post-treatment tool allows to distinguish between the wavepacket LOW and a UP components (the latter being essentially zero in this example). 
The small oscillations in the $ \langle X \rangle$ dynamics are due to the wavepacket initial position that is not exactly the equilibrium position of the adiabatic surface. For the case of the system that is damped by vibrations, the wavepacket can only be represented by the reduced density matrix $\rho_\text{RC}(x,x^\prime)$. The population amplitude in function of the RC value is obtained by taking the diagonal elements of the reduced density matrix. To obtain the UP and LOW parts, the reduced density matrix is projected on the fly aside from the dynamics and results in the reduced density matrix UP $\rho_\text{RC}^{\text{UP}}(x,x^\prime)$ and LOW $\rho_\text{RC}^{\text{LOW}}(x,x^\prime)$. Despite the tunnelling process, the full transition from left to right does not happen here and the system wavepacket loses energy due to vibrations. Therefore, it stabilises towards the ground state which is characterised by the wavepacket uniformly split between the two wells and a mean displacement value $\langle X \rangle = 0~$arb. units.

\begin{figure}
   \includegraphics[scale=1.0]{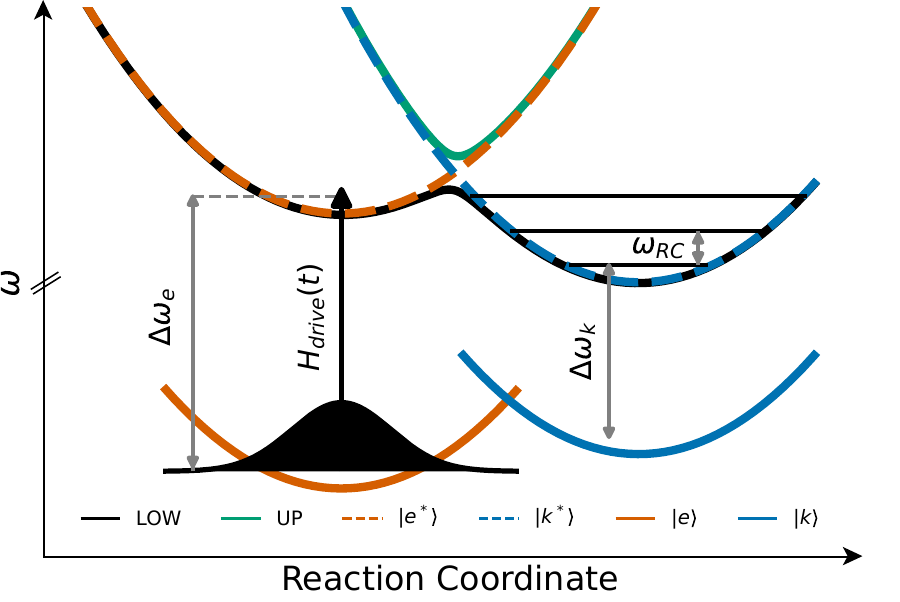}
\caption{Scheme of the energetic landscape of a four-level system described with the RC dimension. Diabatic $\ket{e^*}$ and $\ket{k^*}$ minima are separated by a barrier. The vertical electronic gaps are kept the same as for discretized states. The wavepacket is initialised in $\ket{e}$ and promoted to $\ket{e^*}$ by a pulse $H_\text{drive} (t)$. }
\label{fig:4lvl_RC_adiabdyn_DW}
\end{figure}

\newpage
\subsection{Excited-State Proton Transfer under Driving: Exact Treatment of Dissipative Non-Adiabatic Dynamics}

The second example takes the previous four-level model further by incorporating the RC dimension to it (Figure~\ref{fig:4lvl_RC_adiabdyn_DW}). As in Figure~\ref{fig:DiabAdiab}, the two excited discrete states are directly coupled and become an unique adiabatic surface with its two local minima separated by a barrier. The wavepacket is initialised in the enol ground state and is promoted by a laser pulse of the form :
\begin{equation}
    H_\text{drive}\left( t \right) = \left( \ket{e^*}\bra{e} + \ket{k^*}\bra{k}  + \text{h.c.} \right) \epsilon \exp{\left( \left(t- t_0 \right)^2 / \left( 2 \tau_\text{drive}^2 \right) \right)} \cos\left( \omega_\text{drive} t \right)
\end{equation}
with $t_0$ the centre time of the pulse and $\tau_\text{drive} = \text{FWHM} / 2 \sqrt{2 \log(2)}$ with $\text{FWHM}$ the full width at half maximum. This time-dependent Hamiltonian simulates a laser pulse with a gaussian envelope. The two previous environments (photons and vibrations) are also included in the full wave-function. Parameters are reported in Table S6 of the Supporting Information.

\begin{figure}
   \includegraphics[scale=1.0]{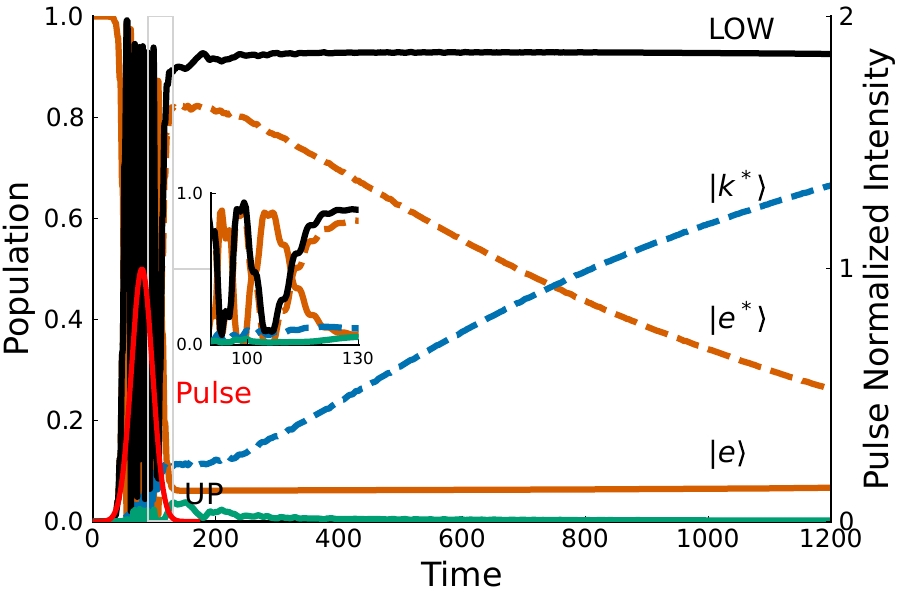}
\caption{Diabatic and excited adiabatic population dynamics. The adiabatic populations are calculated with the adiabatic projector. After being populated by the pulse, coupling with vibrations transfer $\ket{e^*}$ population to $\ket{k^*}$. Time is in arbitrary units.}
\label{fig:4lvl_RC_adiabdyn_pop}
\end{figure}

\begin{figure}
   \includegraphics[scale=1.0]{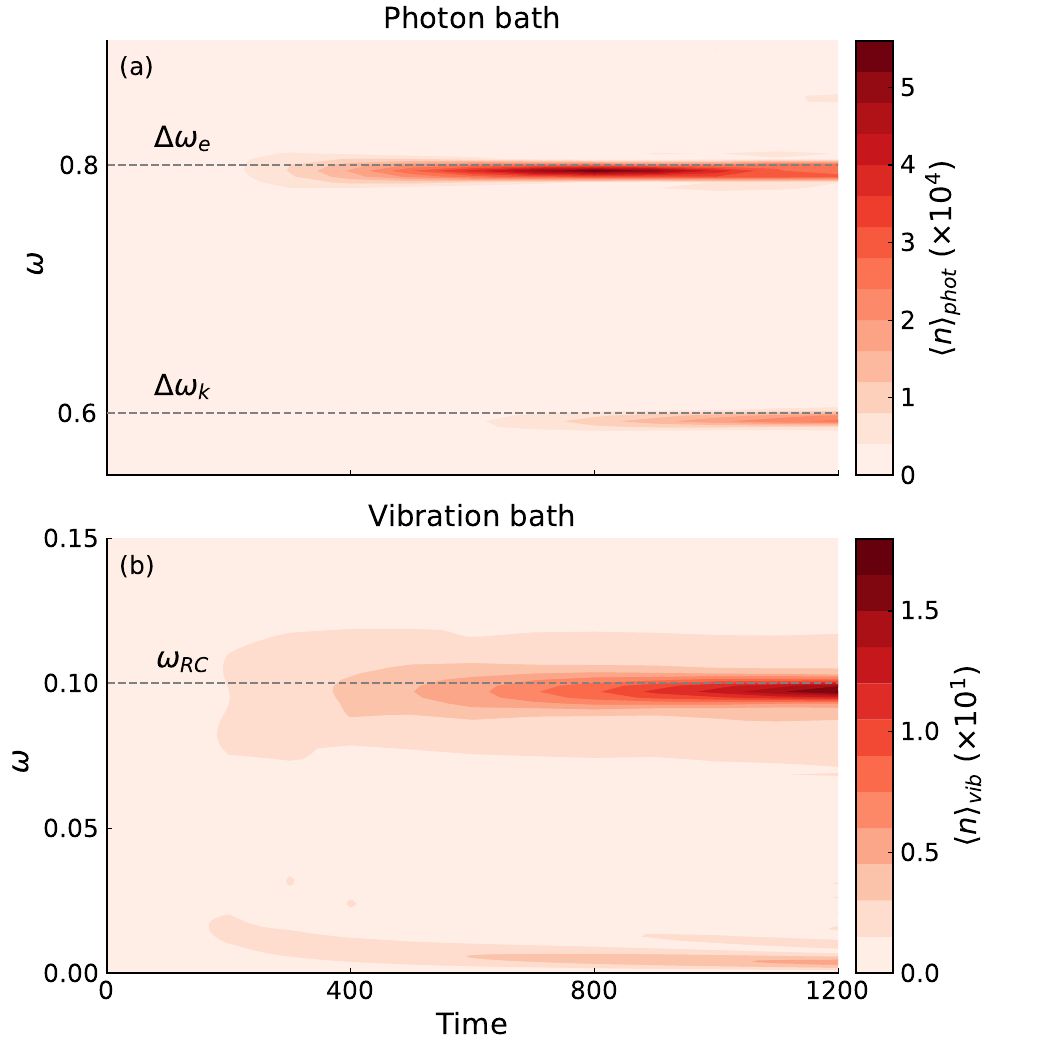}
\caption{(a) Photo-emission spectrum resolved in time. Two peaks are detected at $\Delta \omega_e$ and at $\Delta \omega_k$. (b) Vibration spectrum resolved in time. The main peak is at $\omega_\text{RC}$ but smaller frequencies are also probed. This frequency range is allowed by the anharmonicity of the double-well. Time and frequencies are in arbitrary units.}
\label{fig:4lvl_RC_adiabdyn_spectra}
\end{figure}

From these dynamics, diabatic and adiatic populations are probed. Figure~\ref{fig:4lvl_RC_adiabdyn_pop} shows the initial population in $\ket{e}$ that oscillates due to the pulse. After the pulse, the electronic population is mainly in $\ket{e^*}$ with a little bit of $\ket{k^*}$ and $\ket{e}$. For the rest of the dynamics, $\ket{e}$ population seems constant while a transfer between $\ket{e^*}$ and $\ket{k^*}$ occurs. After the pulse the excited state population is principally in the LOW excited adiabatic state. Two peaks are detected in the time-resolved photon spectrum at frequency $\Delta \omega_e$ and $\Delta \omega_k$ (Figure~\ref{fig:4lvl_RC_adiabdyn_spectra} (a)), whereas that in the vibration spectrum, the main peak is at $\omega_\text{RC}$ but other less energetic excitations can be detected (Figure~\ref{fig:4lvl_RC_adiabdyn_spectra} (b)). 

From the population dynamics, the same behaviour as for discretized states is described. Unlike the previous example of a four-level under strong driving, the gaussian shape of the pulse does not induce dressed states in the dynamics. The pulse populates $\ket{e^*}$ and vibrations induce transition from $\ket{e^*}$ to $\ket{k^*}$, without involving non-adiabatic transitions. The damping induced by both environments will make the wavepacket to fall down in energetic minima of the excited manifold while simultaneously slightly drop down towards ground states by emitting a photon. Although the main vibrational peak is at $\omega_\text{RC}$, several frequencies can be detected in the vibration spectrum. In fact, the coupling between the two wells allows diverse gap energies between vibrational states, broadening the detected vibrational frequency range. It is important to point out that, although the space dimension has been added in this example, the dual fluorescence in the photo-emission spectrum is recovered. This model shows the same main characteristics of the discretized four-level system but with more information concerning the proton wavepacket. \\

At this stage, one can really appreciate the potential of the method and the new features. As a matter of fact, by combining the RC mode with the time-dependent Hamiltonian, the problematic question in the initialisation of the wavepacket is bypassed. This specific issue can represent a bottleneck for the use of numerical method such as Surface Hopping when the Franck-Condon region includes several vibrational modes or in order to include the initial energy distribution \cite{Avagliano2022,Pieroni2023}. \\

\begin{figure}
   \includegraphics[scale=1.0]{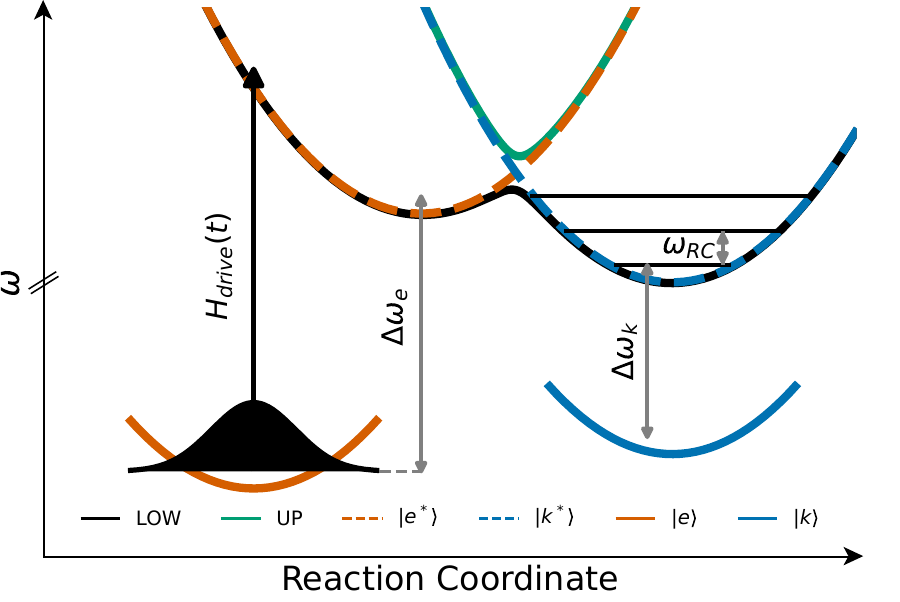}
\caption{Scheme of the energetic landscape of a four-level system described with the RC dimension. The wavepacket is initialised in $\ket{e}$ that is shifted from $\ket{e^*}$ and promoted by a pulse $H_\text{drive} (t)$. The promoted wavepacket has a high energy and undergoes non-adiabatic transitions.}
\label{fig:4lvl_RC_nonadiab_DW}
\end{figure}

Dynamics with different parameters were conducted to illustrate non-adiabatic transitions. For the example showed in Figure~\ref{fig:4lvl_RC_nonadiab_DW}, the system is the same as the previous example except that the enol ground state is shifted from the enol excited state and the driving has a higher frequency, leading to a higher energy for the promoted wavepacket. Parameters are reported in Table S7 of the Supporting Information. \\

\begin{figure}
   \includegraphics[scale=1.0]{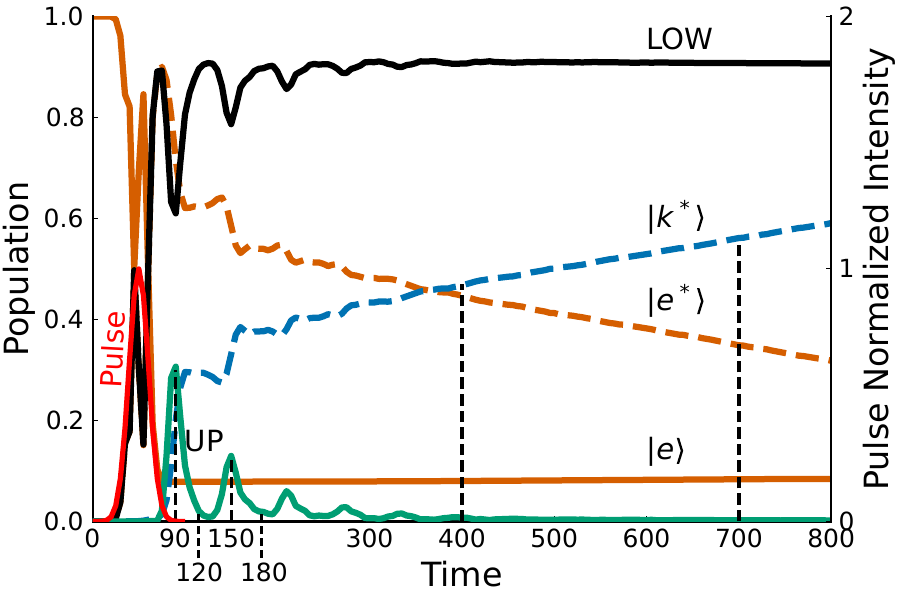}
\caption{Diabatic and excited adiabatic population dynamics. Adiabatic populations are calculated with the adiabatic projector and illustrate non-adiabatic transitions. After being populated by the pulse, $\ket{e^*}$ transfers its population to $\ket{k^*}$ thanks to the coupling with vibrations. Non-adiabatic transitions occur periodically in the short times. Time is in arbitrary units.}
\label{fig:4lvl_RC_nonadiab_pop}
\end{figure}

\begin{figure}
   \includegraphics[scale=1.0]{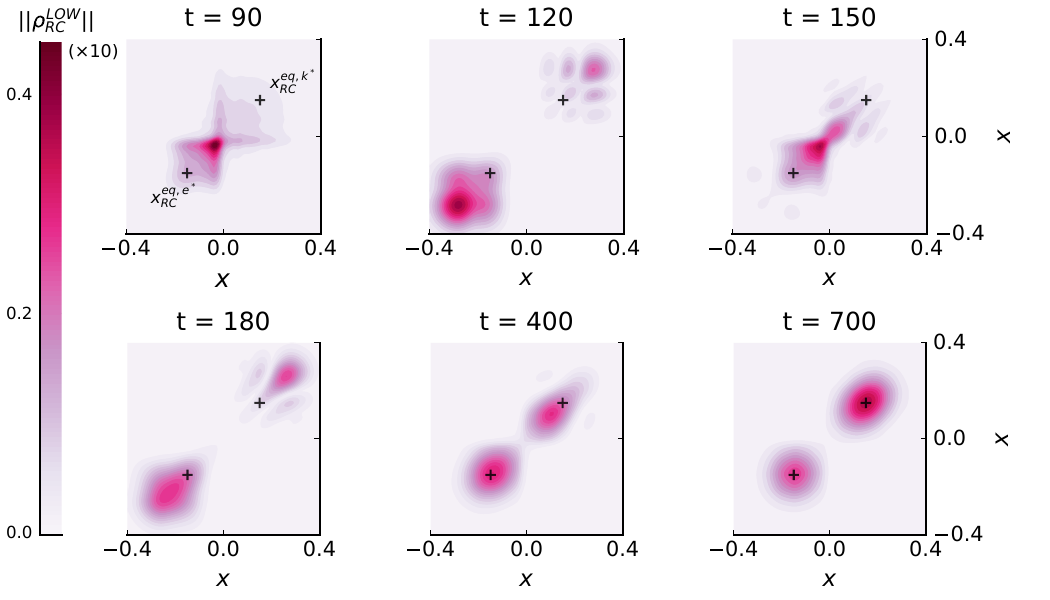}
\caption{Time snapshots of the adiabatic reduced density matrix $|| \rho_\text{RC}^\text{LOW}(x,x^\prime)||$. Equilibrium positions of the diabatic excited states are represented by the black crosses. Non-adiabatic transitions occur when the wavepacket becomes close to the barrier. $x$ and $t$ are in arbitrary units.}
\label{fig:4lvl_RC_nonadiab_rholow}
\end{figure}

The time evolution of the different states is reported in Figure~\ref{fig:4lvl_RC_nonadiab_pop}, showing a behaviour similar to the previous example. The main difference is the UP population that is exhibiting periodical peaks at short times. With the same period, the $\ket{e^*}$ population undergoes a steep decrease and is transferred to $\ket{k^*}$. These adiabatic population variations highlight non-adiabatic transitions, whereas the steps between diabatic population suggest a tunnelling process. Despite non-adiabatic transitions at short times, the excited-state electronic transfer at longer times is similar to the previous example. \newline
In Figure~\ref{fig:4lvl_RC_nonadiab_rholow}, the reduced density matrix is represented and illustrates a useful tool here. This observable in space gives insights into the position onto the excited states. The diagonal elements of this observable represent the population in space whereas anti-diagonal elements represent coherences. The investigation of the adiabatic reduced density matrix for the excited states allows to observe the oscillation of the wavepacket onto the left well, transferring its weight little by little towards the right well. This space monitoring along the RC helps detecting that non-adiabatic transitions happen when the left wavepacket oscillation gets closer to the barrier at time $t=90~$arb. units and $t=150~$arb. units for instance. The frequency of these wavepacket oscillations corresponds to the RC mode frequency $\omega_\text{RC} = 2 \pi/T$ with $T \approx 60~$arb. units. Once vibration dissipation takes effect and stabilises the main wavepacket onto the bottom of the left surface (from $t=400~$arb. units), an adiabatic dynamics is shown for the transfer from $\ket{e^*}$ to $\ket{k^*}$, where the excited keto wavepacket is growing at the expense of the excited enol wavepacket. A last interesting observation concerns the anti-diagonal elements of the reduced density matrix. In fact, the space coherences in $|| \rho_\text{RC}^\text{LOW}(x,x^\prime)||$ are probed especially in short times. These features are characteristic of quantum out-of-equilibrium dynamics and they stay significant until the dissipation due to vibrations kills them. These quantum coherences in space are a result of the Hamiltonian constructed with electronic states and a RC quantum harmonic oscillator.


\newpage

\section{Conclusion}
\label{sec:conclusion}

In this work, we present two models to tackle the problem of proton transfer with environment interactions using Matrix Product State. A first model represents a four-level electronic system under driving that incorporates intramolecular vibration and photon interactions. Secondly, we extended the TEDOPA algorithm to explicitly include a continuous reaction-coordinate within the electronic system, leading to a real-space double-well description of ESIPT which we studied at a quantum level. This description allows to study the electronic transitions as well as the proton displacement on energy landscapes with a full treatment of non-adiabiatic effects. Developing an numerically efficient MPO projector, the resulting dissipative dynamics were analysed in both diabatic and adiabatic bases, demonstrating that the method is an efficient and complete tool to study non-adiabatic processes for open quantum systems. This approach can naturally be extended for further study with thermalised (finite-temperature) vibrations, as well as realistic, structured spectral densities based on \textit{ab initio} calculations \cite{riva2023thermal,hunter2024environmentally}, following the ideas recently presented in Refs.\cite{riva2023thermal,hunter2024environmentally}. Although the energy surfaces presented here were generic toy models, more specific energy landscapes such as a Marcus inverted region, different RC frequencies for $\ket{e^*}$ and $\ket{k^*}$, or multiple surfaces extracted from experimental and/or \textit{ab initio} data could be studied with this methodology, and would provide interesting outlooks. Future work in our team aims to consolidate these developments to model ESIPT in real molecules where ultrafast experimental data is available. 

\begin{suppinfo}

Additional information about the MPO representation of the models. Analytical results and additional observables for the case of the four-level system under strong driving. 

\end{suppinfo}

\begin{acknowledgement}
We thank iSiM (Initiative Sciences et ing\'enierie mol\'eculaires) 
 from the Alliance Sorbonne Universit\'e for funding.

\end{acknowledgement}

\bibliography{biblio}

\end{document}



All the calculations of this work were done with the MPSDynamics \cite{MPSDynamics} Julia package  that can be found along with the documentation at \url{https://github.com/shareloqs/MPSDynamics}.
 
\section{MPO representation with the TEDOPA procedure}

To build the MPO, the Hamiltonian has to be written locally using TEDOPA \cite{Tamascelli2020,Prior2010, chin_exact_2010}. For the first part of the study, one electronic system is interacting with two different environments. The Hamiltonian reads:
\begin{subequations}
\renewcommand{\theequation}{\theparentequation.\arabic{equation}}
\begin{align} \label{eq:4lvl}
&H\left( t \right) = H_\text{S}  + H_\text{drive}\left( t \right) + H_\text{B} + H_\text{SB} \\
    &H_\text{S} = \omega_{e} \ket{e}\bra{e} + \omega_{e^*} \ket{e^*}\bra{e^*} + \omega_{k} \ket{k}\bra{k} + \omega_{k^*} \ket{k^*}\bra{k^*} \\
    &H_\text{drive}\left( t \right) = \left( \ket{e^*}\bra{e} + \ket{k^*}\bra{k}  + \text{h.c.} \right) \epsilon \sin\left( \omega_\text{drive} t \right) \\
    &H_\text{B} = \overbrace{\sum_k \omega_k^\text{phot} \left(a_k^{\dagger}a_k\right)}^{\text{Photons}} + \overbrace{\sum_{k^\prime} \omega_{k^\prime}^\text{vib} \left(b_{k^\prime}^{\dagger}b_{k^\prime}\right)}^{\text{Vibrations}}  \\
    &H_\text{SB} =  \left(\ket{e^*}\bra{e} + \ket{k^*}\bra{k} + \text{h.c} \right) \sum_k g_k \left(a_k^{\dagger} + a_k \right)  + \left(\ket{e^*}\bra{k^*} + \text{h.c} \right) \sum_{k^\prime} y_{k^\prime} \left( b_{k^\prime}^{\dagger} +  b_{k^\prime} \right)
\end{align}
\end{subequations}

These environment are linearized thanks to the TEDOPA procedure and results in the main system interacting with only the first mode of the chains. In order to write the MPO (eq \ref{eq:4lvl}), several tensors have to express the Hamiltonian locally. The MPO has the following form: 
\begin{equation} \label{eq:MPO_4lvl}
         H = W^{\text{phot}}_n \otimes \ldots \otimes W^{\text{phot}}_2 \otimes W^{\text{phot}}_{1} \otimes W_\text{S}  \otimes W^{\text{vib}}_1 \otimes W^{\text{vib}}_2 \otimes \ldots \otimes W^{\text{vib}}_m
\end{equation}

The electronic tensor is a translation of the electronic part of eq \ref{eq:4lvl} ($H_\text{S} + H_\text{S} \left( t \right)$) adding the photon and vibration first chain mode interaction:
\begin{equation}\label{eq:elec_tensor}
    W_\text{S} =  \begin{pmatrix}
      H_{\text{S}} + H_{\text{drive}} \left( t \right)& t_0^\text{vib} \left(\ket{e^*}\bra{k^*} + \text{h.c} \right)  &  \mathbb{1}  \\
        t_0^\text{phot} \left(\ket{e^*}\bra{e} + \ket{k^*}\bra{k} + \text{h.c} \right) & 0 &  0  \\
        \mathbb{1} & 0  &  0 \\
    \end{pmatrix}
\end{equation}

where $t_0^\text{phot}$ and $t_0^\text{vib}$ represent the coupling coefficients between the system and the first mode of the chain of photon and vibration respectively. Next, the photon and vibration chain can be written with tensors for a mode $i$ following the usual TEDOPA structure and the MPSDynamics package formalism:
\begin{equation} \label{MPO_chains}
    W^\text{phot}_i =  \begin{pmatrix}
       \mathbb{1} &  \left( \tilde{a_i} + \tilde{a_i}^\dagger \right) & \epsilon_i^\text{phot} \left( \tilde{a_i}^\dagger \tilde{a_i}\right) \\
       0 & 0 & t_i^\text{phot} \tilde{a_i} \\
       0 & 0 & t_i^\text{phot} \tilde{a_i}^\dagger \\
       0 & 0 & \mathbb{1}
    \end{pmatrix} \, \, \, \, \, W^\text{vib}_i =  \begin{pmatrix}
       \mathbb{1} & 0  &  0 &  0 \\
        \left( \tilde{b_i} + \tilde{b_i}^\dagger \right) & 0 & 0  & 0 \\
        \epsilon_i^\text{vib}\left( \tilde{b_i}^\dagger \tilde{b_i}\right) & t_i^\text{vib} \tilde{b_i}&  t_i^\text{vib} \tilde{b_i}^\dagger & \mathbb{1}
    \end{pmatrix}
\end{equation}

with $\tilde{a_i}^{\left(\dagger\right)}$ the annihilation (creation) operator of the $i^\text{th}$ mode of the photon chain. This mode is characterised by its energy $\epsilon_i^\text{phot}$ and first-neighbour coefficient between the $i^\text{th}$ and the $\left(i+1\right)^\text{th}$ modes $t_i^\text{phot}$. $\tilde{b_i}^{\left(\dagger\right)}$ is the annihilation (creation) operator of the $i^\text{th}$ mode of the vibration chain that is described in the same manner, with the coefficients $\epsilon_i^\text{vib}$ and $t_i^\text{vib}$. These coefficients and operators are obtained with the help of TEDOPA. We refer the reader to other works detailing the theory and how to obtain these coefficients \cite{chin_exact_2010}. \\

When multiplying all of these bounded tensors, one recovers the full Hamiltonian in the linearised basis. One has to keep in mind that each term of the represented matrices $W$ is itself a matrix. Environments and system formulations can be modified as desired, emphasising the versatility of the presented method to calculate different dynamics. \\


Next, including the Reaction Coordinate, the Hamiltonian reads:
\begin{subequations}
\renewcommand{\theequation}{\theparentequation.\arabic{equation}}
\begin{align} \label{eq:RC_wo_bath}
    &H = H_{\text{S}} + + H_\text{drive}\left( t \right) + H_{\text{RC}} + H_{\text{int}}^{\text{RC-S}} + H_{\text{vib}} + H_{\text{int}}^{\text{vib-RC}}\\ \label{eq:TLS_RC}
    &H_{\text{S}} = \omega_{e^*}\ket{e^*}\bra{e^*} + \omega_{k^*}\ket{k^*}\bra{k^*} + \Delta \left( \ket{e^*}\bra{k^*} + \ket{k^*}\bra{e^*} \right) \\ 
    &H_{\text{RC}} = \omega_{\text{RC}} \left(d^{\dagger}d + \frac{1}{2} \right)\\
    &H_{\text{int}}^{\text{RC-S}} = g_{e^*} \ket{e^*}\bra{e^*}\left( d + d^{\dagger}\right)+ g_{k^*} \ket{k^*}\bra{k^*}\left( d + d^{\dagger}\right) \\
        &H_{\text{vib}} = \sum_k \omega_k b_k^{\dagger} b_k \\
      &H_{\text{int}}^{\text{vib-RC}} = \lambda_{\text{reorg}} \left(d + d^{\dagger} \right)^2 - \left(d + d^{\dagger} \right) \sum_k z_k \left(b_k + b_k^{\dagger} \right)  \label{eq:RC_vibcoupl2}
\end{align}
\end{subequations}

To translate this Hamiltonian into an MPO, the main system consists in two tensors. One of the tensor characterises the electronic system and the other tensor defines a quantum harmonic oscillator. Photons are now coupled to electrons and vibrations are coupled to the RC mode. The MPO is then expressed in the shape described in eq \ref{eq:MPO_RC}.
\begin{equation} \label{eq:MPO_RC}
         H = W^{\text{phot}}_n \otimes \ldots \otimes W^{\text{phot}}_2 \otimes W^{\text{phot}}_{1} \otimes W_\text{S} \otimes W_\text{RC} \otimes W^{\text{vib}}_1 \otimes W^{\text{vib}}_2 \otimes \ldots \otimes W^{\text{vib}}_m
\end{equation}

Concerning the system tensor, which specifies the electronic Hamiltonian ($H_\text{S} + H_\text{S} \left(t \right)$) , the coupling with the RC mode ($H_\text{int}^\text{RC-S}$) and  the photon interaction, it reads:
\begin{equation}
    W_\text{S} =  \begin{pmatrix}
      H_{\text{S}} + H_{\text{drive}} \left( t \right) & g_{e^*} \ket{e^*}\bra{e^*} +g_{k^*} \ket{k^*}\bra{k^*}  &  \mathbb{1}  \\
        t_0^\text{phot} \left( \ket{e^*}\bra{k^*} + \ket{k^*}\bra{e^*} \right) & 0 &  0  \\
        \mathbb{1} & 0  &  0 \\
    \end{pmatrix}
\end{equation}

Eventually, taking into account eq \ref{eq:RC_vibcoupl2}, the reaction coordinate tensor is written as:
\begin{equation}
    W_\text{RC} =  \begin{pmatrix}
       \mathbb{1} & 0  &  0  \\
        \left(d + d^{\dagger}\right)  & 0 & 0   \\
        \omega_\text{RC} \left(d^{\dagger}d + \frac{1}{2} \right)+ \lambda_\text{reorg} \left(d +d^{\dagger} \right)^2 & - t_0^\text{vib}\left(d+d^{\dagger} \right)  & \mathbb{1}  \\
    \end{pmatrix}
\end{equation}
It has to be kept in mind that the initial MPS is described as a displaced state for the RC mode, allowing the wavepacket to be initialised at a certain RC value. 

\section{Analytical results for the strong driving case}

In order to evaluate the dressed state energies, the new Hamiltonian matrix has to be diagonalized. The driving term leads to an interaction between a state $\ket{e^*,n}$ and $\ket{e,n+1}$ with $n$ the number of absorbed photon. Because the driving is on resonance, these states are degenerate at energy $\omega_{e^*,n} = \omega_{e^*} + n \times \omega_\text{drive} =  \omega_{e} + \left( n + 1\right) \omega_\text{drive} = \omega_{e,n+1}$. The following matrix reads:

\begin{equation}
        H_\text{S}^n =  \begin{pmatrix}
      \omega_{e^*,n} & \epsilon /2 \\
       \epsilon /2  &  \omega_{e,n+1}\\
    \end{pmatrix}
\end{equation}

The new eigenvalues for the excited enol can be simplified as:
\begin{align}
    \tilde{\omega}_{e^*,n} &= \omega_{e^*,n} \pm \frac{1}{2} \, \sqrt{4 \left( \frac{\epsilon}{2} \right)^2} \\
    \tilde{\omega}_{e^*,n} &= \omega_{e^*,n} \pm \frac{\epsilon}{2}
\end{align}
The same is true for the enol ground state ( $\tilde{\omega}_{e,n} = \omega_{e,n} \pm \frac{\epsilon}{2}$). Concerning the keto states, the drive is no longer on resonance. The excited keto state $\ket{k^*,n}$ at energy $\omega_{k^*} + n \times \omega_\text{drive}$ is interacting with the $\ket{k,n+1}$ state at energy $\omega_{k} + \left( n +1 \right) \times \omega_\text{drive}$. Taking the example of $n=0$, the diagonalization leads to :
\begin{align}
    \tilde{\omega}_{k^*,n} = \frac{1}{2} \left( \omega_{k^*,n} + \omega_{k,n+1} \right)  \pm \frac{1}{2} \sqrt{\left( \omega_{k,n+1} - \omega_{k^*,n} \right)^2 + 4 \left( \frac{\epsilon}{2} \right)^2}
\end{align}
Being off resonance, the state that has the largest electronic contribution state is the higher in energy for $\ket{k}$ and the lower in energy for $\ket{k^*}$, such that

\begin{align}
    \tilde{\omega}_{k^*, n} &= \frac{1}{2} \left( \omega_{k^*, n} +  \omega_{k, n+1} \right)  - \frac{1}{2} \sqrt{\left( \omega_{k, n+1} -\omega_{k^*, n}  \right)^2 + 4 \left( \frac{\epsilon}{2} \right)^2} \\
        \tilde{\omega}_{k, n+1} &= \frac{1}{2} \left( \omega_{k^*, n} +  \omega_{k, n+1} \right)  + \frac{1}{2} \sqrt{\left( \omega_{k, n+1} -\omega_{k^*,n}  \right)^2 + 4 \left( \frac{\epsilon}{2} \right)^2}
\end{align}
which leads to a new keto gap that reads
\begin{align}
    \Delta \tilde{\omega}_k &= \tilde{\omega}_{k^*,n} - \tilde{\omega}_{k,n} \\
    \Delta \tilde{\omega}_k &= \frac{1}{2} \left( \omega_{k^*, n} +  \omega_{k, n+1} - \omega_{k^*, n-1} - \omega_{k, n}\right) - \frac{1}{2} \sqrt{\left( \omega_{k, n+1} -\omega_{k^*,n}\right)^2 + \epsilon^2 } - \frac{1}{2} \sqrt{\left( \omega_{k, n} -\omega_{k^*,n-1}\right)^2 + \epsilon^2 } 
\end{align}

\section{Supplementary figures for the strong driving case}

\begin{figure}
    \centering
    \includegraphics[scale=1.0]{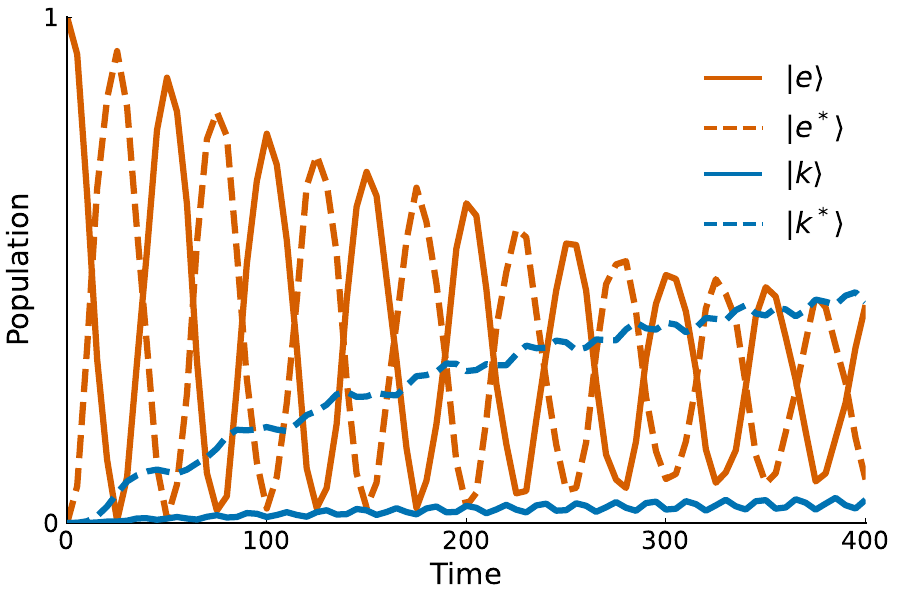}
    \caption{Electronic population of a strongly-driven four-level system coupled to two environments. Time is in arbitrary units.}
    \label{fig:pop_strongdriving}
\end{figure}

The electronic populations of the strong-driving case are shown in Figure \ref{fig:pop_strongdriving}.

\begin{figure}
    \centering
    \includegraphics[scale=1.0]{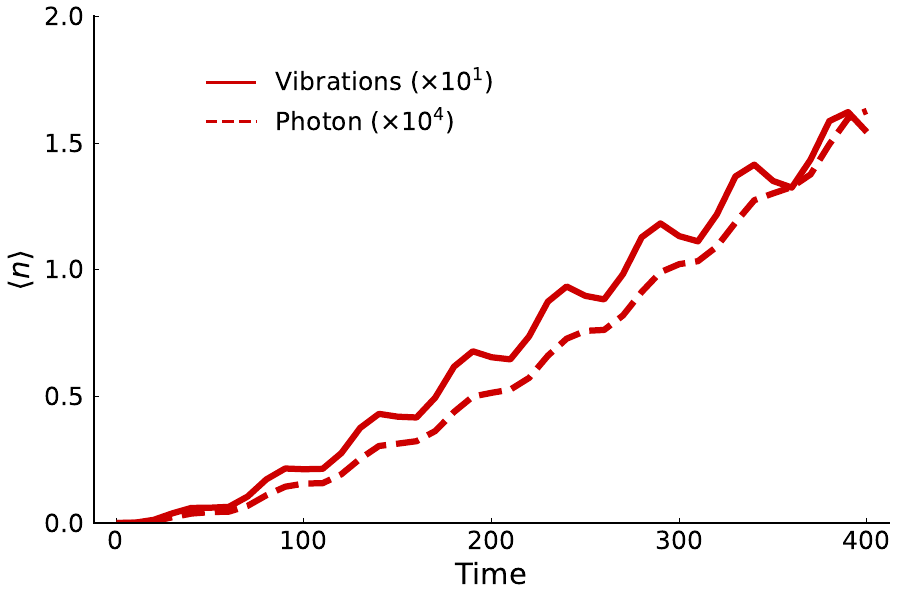}
    \caption{Population averaged of vibrations at frequency $\omega_{e^*} - \omega_{k^*}$ and photons at frequency $\Delta \omega_e$ are represented in time for the strong driving case. Oscillations are due to the applied field. Time is in arbitrary units.}
    \label{fig:osci_strongdriving}
\end{figure}

Figure \ref{fig:osci_strongdriving} shows the oscillations of the highest emitted peak in the strong driving case. These oscillations in the environments are due to the driving, impacting the emission in time. 

\subsection{Parameters}
We present here the parameters for the different examples. Regarding the environment characteristics, we used continuous spectral densities of the form  $J(\omega) = 2 \alpha \, \omega^s/\omega_c^{s-1}$ with a strong cutoff at $\omega_c$. These spectral densities are used for the TEDOPA procedure. 
\begin{table} 
\caption{Four-level system, no driving case. All of the values are in arbitrary units.}

\begin{minipage}{0.45\textwidth}
\centering
\begin{tabular}{|c|c|}
\hline
\textbf{Parameter} & \textbf{Value} \\
 \hline  $\omega_e$ & $-0.4$ \\
\hline $\omega_{e^*}$ & $0.4$\\
\hline $\omega_k$ & $-0.3$\\
\hline $\omega_{k^*}$ & $0.3$\\
\hline $dt$ & $1.0$\\
\hline $t_\text{final}$ & $800.0$\\
\hline $D_{\text{max}}$ & $6$\\
\hline $\epsilon_{\text{drive}}$ & $0$\\
\bottomrule
\end{tabular}

\end{minipage}
\begin{minipage}{0.45\textwidth}
\centering
\begin{tabular}{|c|c|}
\hline
\textbf{Parameter} & \textbf{Value} \\
\hline $\alpha_{\text{phot}}$ & $8.10^{-7}$\\
\hline $N_{\text{phot}}$ & $320$\\
\hline $d_{\text{phot}}$ & $2$\\
\hline $ s_{\text{phot}}$ & $2$\\
\hline $\omega_c^\text{phot}$ & $1.0$\\
\hline $\alpha_{\text{vib}}$ & $8.10^{-3}$\\
\hline $N_{\text{vib}}$ & $120$\\
\hline $d_{\text{vib}}$ & $6$\\
\hline $s_{\text{vib}}$ & $2$\\
\hline $\omega_c^\text{vib}$ & $0.3$\\
\hline
\end{tabular}

\end{minipage}
\end{table}

\begin{table} 
\caption{Four-level system, weak driving case. All of the values are in arbitrary units.}

\begin{minipage}{0.45\textwidth}
\centering
\begin{tabular}{|c|c|}
\hline
\textbf{Parameter} & \textbf{Value} \\
 \hline  $\omega_e$ & $-0.4$ \\
\hline $\omega_{e^*}$ & $0.4$\\
\hline $\omega_k$ & $-0.3$\\
\hline $\omega_{k^*}$ & $0.3$\\
\hline $dt$ & $0.1$\\
\hline $t_\text{final}$ & $1600.0$\\
\hline $D_{\text{max}}$ & $6$\\
\hline $\epsilon_{\text{drive}}$ & $2 \pi / 800$\\
\hline $\omega_{\text{drive}}$ & $\omega_{\text{enol}}$\\
\bottomrule
\end{tabular}

\end{minipage}
\begin{minipage}{0.45\textwidth}
\centering
\begin{tabular}{|c|c|}
\hline
\textbf{Parameter} & \textbf{Value} \\
\hline $\alpha_{\text{phot}}$ & $8.10^{-7}$\\
\hline $N_{\text{phot}}$ & $600$\\
\hline $d_{\text{phot}}$ & $2$\\
\hline $ s_{\text{phot}}$ & $2$\\
\hline $\omega_c^\text{phot}$ & $1.0$\\
\hline $\alpha_{\text{vib}}$ & $8.10^{-3}$\\
\hline $N_{\text{vib}}$ & $240$\\
\hline $d_{\text{vib}}$ & $8$\\
\hline $s_{\text{vib}}$ & $2$\\
\hline $\omega_c^\text{vib}$ & $0.3$\\
\hline
\end{tabular}

\end{minipage}
\end{table}

\begin{table} 
\caption{Four-level system, strong driving case. All of the values are in arbitrary units.}

\begin{minipage}{0.45\textwidth}
\centering
\begin{tabular}{|c|c|}
\hline
\textbf{Parameter} & \textbf{Value} \\
 \hline  $\omega_e$ & $-0.4$ \\
\hline $\omega_{e^*}$ & $0.4$\\
\hline $\omega_k$ & $-0.3$\\
\hline $\omega_{k^*}$ & $0.3$\\
\hline $dt$ & $0.1$\\
\hline $t_\text{final}$ & $400.0$\\
\hline $D_{\text{max}}$ & $6$\\
\hline $\epsilon_{\text{drive}}$ & $2 \pi /50$\\
\hline $\omega_{\text{drive}}$ & $\omega_{\text{enol}}$\\
\bottomrule
\end{tabular}

\end{minipage}
\begin{minipage}{0.45\textwidth}
\centering
\begin{tabular}{|c|c|}
\hline
\textbf{Parameter} & \textbf{Value} \\
\hline $\alpha_{\text{phot}}$ & $8.10^{-7}$\\
\hline $N_{\text{phot}}$ & $200$\\
\hline $d_{\text{phot}}$ & $3$\\
\hline $ s_{\text{phot}}$ & $2$\\
\hline $\omega_c^\text{phot}$ & $1.1$\\
\hline $\alpha_{\text{vib}}$ & $8.10^{-3}$\\
\hline $N_{\text{vib}}$ & $60$\\
\hline $d_{\text{vib}}$ & $8$\\
\hline $s_{\text{vib}}$ & $2$\\
\hline $\omega_c^\text{vib}$ & $0.3$\\
\hline
\end{tabular}

\end{minipage}
\end{table}

\begin{table} 
\caption{Reaction Coordinate system, symmetric doublewell case, no environment. The initial state is projected onto the lower surface before the dynamics. All of the values are in arbitrary units.}

\begin{minipage}{0.45\textwidth}
\centering
\begin{tabular}{|c|c|}
\hline
\textbf{Parameter} & \textbf{Value} \\
\hline $\omega_{e^*}$ & $0.8$\\
\hline $\omega_{k^*}$ & $0.8$\\
\hline $\Delta$ & $0.05$\\
\hline $dt$ & $1.0$\\
\hline $t_\text{final}$ & $3100.0$\\
\hline $D_{\text{max}}$ & $2$\\
\hline $x$ & $(-1.4,1.4)$\\
\hline $\delta x$ & $0.001$\\
\hline $ m$ & $1.83 \times 10^3$\\
\hline $\omega_\text{RC}$ & $0.0347$\\
\hline $x_\text{RC}^{\text{ini}}$ & $-0.25$\\
\hline $x_\text{RC}^{\text{eq,}e^*}$ & $-0.25$\\
\hline $x_\text{RC}^{\text{eq,}k^*}$ & $0.25$\\
\hline $d_\text{Fock}^\text{RC}$ & $45$\\
\bottomrule
\end{tabular}

\end{minipage}

\end{table}

\begin{table} 
\caption{Reaction Coordinate system, symmetric doublewell case, with environment. The initial state is projected onto the lower surface before the dynamics. All of the values are in arbitrary units.}

\begin{minipage}{0.45\textwidth}
\centering
\begin{tabular}{|c|c|}
\hline
\textbf{Parameter} & \textbf{Value} \\
\hline $\omega_{e^*}$ & $0.8$\\
\hline $\omega_{k^*}$ & $0.8$\\
\hline $\Delta$ & $0.05$\\
\hline $dt$ & $1.0$\\
\hline $t_\text{final}$ & $3100.0$\\
\hline $D_{\text{max}}$ & $10$\\
\hline $x$ & $(-1.4,1.4)$\\
\hline $\delta x$ & $0.001$\\
\hline $ m$ & $1.83 \times 10^3$\\
\hline $\omega_\text{RC}$ & $0.0347$\\
\hline $x_\text{RC}^{\text{ini}}$ & $-0.25$\\
\hline $x_\text{RC}^{\text{eq,}e^*}$ & $-0.25$\\
\hline $x_\text{RC}^{\text{eq,}k^*}$ & $0.25$\\
\hline $d_\text{Fock}^\text{RC}$ & $45$\\
\bottomrule
\end{tabular}
\end{minipage}
\begin{minipage}{0.45\textwidth}
\centering
\begin{tabular}{|c|c|}
\hline
\textbf{Parameter} & \textbf{Value} \\
\hline $\alpha_{\text{vib}}$ & $8.10^{-3}$\\
\hline $N_{\text{vib}}$ & $150$\\
\hline $d_{\text{vib}}$ & $10$\\
\hline $s_{\text{vib}}$ & $1$\\
\hline $\omega_c^\text{vib}$ & $5 \times \omega_\text{RC}$\\
\hline
\end{tabular}

\end{minipage}

\end{table}

\begin{table} 
\caption{Reaction Coordinate system, ESIPT example with no shift of $\ket{e}$ state. All of the values are in arbitrary units.}

\begin{minipage}{0.45\textwidth}
\centering
\begin{tabular}{|c|c|}
\hline
\textbf{Parameter} & \textbf{Value} \\
 \hline  $\omega_e$ & $-0.4$ \\
\hline $\omega_{e^*}$ & $0.4$\\
\hline $\omega_k$ & $-0.3$\\
\hline $\omega_{k^*}$ & $0.3$\\
\hline $\Delta$ & $0.05$\\
\hline $dt$ & $1.0$\\
\hline $t_\text{final}$ & $1200.0$\\
\hline $D_{\text{max}}$ & $8$\\
\hline $x$ & $(-0.65,0.65)$\\
\hline $\delta x$ & $0.001$\\
\hline $ m$ & $1.83 \times 10^3$\\
\hline $\omega_\text{RC}$ & $0.1$\\
\hline $x_\text{RC}^{\text{ini}}$ & $-0.15$\\
\hline $x_\text{RC}^{\text{eq,}e}$ & $-0.15$\\
\hline $x_\text{RC}^{\text{eq,}e^*}$ & $-0.15$\\
\hline $x_\text{RC}^{\text{eq,}k^*}$ & $0.15$\\
\hline $d_\text{Fock}^\text{RC}$ & $25$\\
\bottomrule
\end{tabular}
\end{minipage}
\begin{minipage}{0.45\textwidth}
\centering
\begin{tabular}{|c|c|}
\hline
\textbf{Parameter} & \textbf{Value} \\
\hline $\alpha_{\text{phot}}$ & $8.10^{-7}$\\
\hline $N_{\text{phot}}$ & $540$\\
\hline $d_{\text{phot}}$ & $2$\\
\hline $ s_{\text{phot}}$ & $2$\\
\hline $\omega_c^\text{phot}$ & $1.6$\\
\hline $\alpha_{\text{vib}}$ & $5.10^{-3}$\\
\hline $N_{\text{vib}}$ & $330$\\
\hline $d_{\text{vib}}$ & $10$\\
\hline $s_{\text{vib}}$ & $1$\\
\hline $\omega_c^\text{vib}$ & $0.575$\\
\hline $\epsilon_{\text{drive}}$ & $1.0$\\
\hline $\omega_{\text{drive}}$ & $0.82$\\
\hline FWHM & $40$\\
\hline $t_0^\text{drive}$ & $80$\\
\hline
\end{tabular}

\end{minipage}

\end{table}

\begin{table} 
\caption{Reaction Coordinate system, ESIPT example with shift of $\ket{e}$ state. The effect of photons being negligible, they were ignored in this example. All of the values are in arbitrary units.}

\begin{minipage}{0.45\textwidth}
\centering
\begin{tabular}{|c|c|}
\hline
\textbf{Parameter} & \textbf{Value} \\
 \hline  $\omega_e$ & $-0.4$ \\
\hline $\omega_{e^*}$ & $0.4$\\
\hline $\omega_k$ & $-0.3$\\
\hline $\omega_{k^*}$ & $0.3$\\
\hline $\Delta$ & $0.05$\\
\hline $dt$ & $0.5$\\
\hline $t_\text{final}$ & $800.0$\\
\hline $D_{\text{max}}$ & $10$\\
\hline $x$ & $(-0.85,0.85)$\\
\hline $\delta x$ & $0.001$\\
\hline $ m$ & $1.83 \times 10^3$\\
\hline $\omega_\text{RC}$ & $0.1$\\
\hline $x_\text{RC}^{\text{ini}}$ & $-0.45$\\
\hline $x_\text{RC}^{\text{eq,}e}$ & $-0.45$\\
\hline $x_\text{RC}^{\text{eq,}e^*}$ & $-0.15$\\
\hline $x_\text{RC}^{\text{eq,}k^*}$ & $0.15$\\
\hline $d_\text{Fock}^\text{RC}$ & $45$\\
\bottomrule
\end{tabular}
\end{minipage}
\begin{minipage}{0.45\textwidth}
\centering
\begin{tabular}{|c|c|}
\hline
\textbf{Parameter} & \textbf{Value} \\
\hline $\alpha_{\text{vib}}$ & $5.10^{-3}$\\
\hline $N_{\text{vib}}$ & $200$\\
\hline $d_{\text{vib}}$ & $10$\\
\hline $s_{\text{vib}}$ & $1$\\
\hline $\omega_c^\text{vib}$ & $0.575$\\
\hline $\epsilon_{\text{drive}}$ & $1.0$\\
\hline $\omega_{\text{drive}}$ & $1.1235$\\
\hline FWHM & $25$\\
\hline $t_0^\text{drive}$ & $50$\\
\hline
\end{tabular}

\end{minipage}

\end{table}

  \newpage
  
\bibliography{biblio}